\documentclass{emulateapj}
\usepackage{graphicx}
\usepackage{mathrsfs}	
\usepackage{amsmath}
\usepackage{color}
\usepackage{hyperref}

\newcommand{\Survey}{BELLS GALLERY}

\shorttitle{BELLS III.}
\shortauthors{Shu et al. 2016}
 
\begin{document}
 
\title{The BOSS Emission-Line Lens Survey. III. : \\
Strong Lensing of L\MakeLowercase{y}$\alpha$ Emitters by Individual Galaxies}

\author{\mbox{Yiping Shu\altaffilmark{1, 2}}}
\author{\mbox{Adam S. Bolton\altaffilmark{2, 3}}}
\author{\mbox{Christopher S. Kochanek\altaffilmark{4}}}
\author{\mbox{Masamune Oguri\altaffilmark{5, 6, 7}}}
\author{\mbox{Ismael P\'{e}rez-Fournon\altaffilmark{8, 9}}}
\author{\mbox{Zheng Zheng\altaffilmark{2}}}
\author{\mbox{Shude Mao\altaffilmark{1, 10, 11}}}
\author{\mbox{Antonio D. Montero-Dorta\altaffilmark{2}}}
\author{\mbox{Joel R. Brownstein\altaffilmark{2}}}
\author{\mbox{Rui Marques-Chaves\altaffilmark{8, 9}}}
\author{\mbox{Brice M\'{e}nard\altaffilmark{12}}}

\altaffiltext{1}{National Astronomical Observatories, Chinese Academy of Sciences, 20A Datun Road, Chaoyang District, Beijing 100012, China ({\tt yiping.shu@nao.cas.cn})}
\altaffiltext{2}{Department of Physics and Astronomy, University of Utah,
115 South 1400 East, Salt Lake City, UT 84112, USA}
\altaffiltext{3}{National Optical Astronomy Observatory, 950 N. Cherry Ave., Tucson, AZ 85719 USA ({\tt bolton@noao.edu})}
\altaffiltext{4}{Department of Astronomy \& Center for Cosmology and Astroparticle Physics, Ohio State University, Columbus, OH 43210, USA}
\altaffiltext{5}{Research Center for the Early Universe, University of Tokyo, 7-3-1 Hongo, Bunkyo-ku, Tokyo 113-0033, Japan}
\altaffiltext{6}{Department of Physics, University of Tokyo, 7-3-1 Hongo, Bunkyo-ku, Tokyo 113-0033, Japan}
\altaffiltext{7}{Kavli Institute for the Physics and Mathematics of the Universe (Kavli IPMU, WPI), University of Tokyo, Chiba 277-8583, Japan}
\altaffiltext{8}{Instituto de Astrof\'{i}sica de Canarias, C/ V\'{i}a L\'{a}ctea, s/n, 38205 San Crist\'{o}bal de La Laguna, Tenerife, Spain}
\altaffiltext{9}{Universidad de La Laguna, Dpto. Astrof\'{i}sica, E-38206 La Laguna, Tenerife, Spain}
\altaffiltext{10}{Physics Department and Tsinghua Centre for Astrophysics, Tsinghua University, Beijing 100084, China}
\altaffiltext{11}{Jodrell Bank Centre for Astrophysics, School of Physics and Astronomy, The University of Manchester, Oxford Road, Manchester M13 9PL, UK}
\altaffiltext{12}{Department of Physics and Astronomy, Johns Hopkins University, Baltimore, MD 21218, USA}

\begin{abstract}

We introduce the Baryon Oscillation Spectroscopic Survey (BOSS) Emission-Line Lens 
Survey (BELLS) for GALaxy-Ly$\alpha$ EmitteR sYstems (\Survey{}) Survey,  
which is a Hubble Space Telescope program to image
a sample of galaxy-scale strong gravitational lens candidate systems with 
high-redshift Ly$\alpha$ emitters (LAEs) as the background sources. 
The goal of the \Survey{} Survey is to illuminate dark substructures 
in galaxy-scale halos by exploiting the small-scale clumpiness of rest-frame
far-UV emission in lensed LAEs, and to thereby constrain the slope and
normalization of the substructure-mass function. 
In this paper, we describe in detail the spectroscopic strong-lens selection 
technique, which is based on methods adopted in the previous Sloan Lens ACS 
(SLACS) Survey, BELLS, and SLACS for the Masses Survey. 
We present the \Survey{} sample of the 21 highest-quality galaxy--LAE candidates 
selected from $\approx 1.4 \times 10^6$ galaxy spectra in the 
BOSS of the Sloan Digital Sky Survey III. These systems consist of massive 
galaxies at redshifts of approximately 0.5 strongly lensing LAEs at redshifts 
from 2--3. The compact nature of LAEs makes them an ideal probe of 
dark substructures, with a substructure-mass sensitivity that is unprecedented in
other optical strong-lens samples.
The magnification effect from lensing will also reveal the 
structure of LAEs below 100 pc scales, providing a detailed look at the 
sites of the most concentrated unobscured star formation in the universe. 
The source code used for candidate selection is available for download as a part of 
this release. 

\end{abstract}


\slugcomment{Submitted to the ApJ}

\maketitle

\section{Introduction}

To date, various observational discoveries including the accelerating expansion of 
the universe, primordial nucleosynthesis, the structure of the 
cosmic microwave background, and the large-scale clustering of galaxies
support the $\Lambda$ cold dark matter ($\rm \Lambda CDM$) model as 
the standard cosmological paradigm 
\citep[e.g.,][]{Riess98, Perlmutter99, Burles01, Anderson14, Planck15}. 
Nevertheless, the nature of dark matter (DM) itself is still mysterious. 
In particular, analytical calculations and numerical simulations based on 
the CDM assumption suggest a significant excess in the number of substructures 
as compared to the observed population of satellite galaxies around the Milky Way 
\citep[e.g.,][]{Klypin99, Moore99a, Bullock10, BK11}. Hence, observationally 
quantifying the abundance and mass function of substructure, especially 
\emph{dark} substructure, represents an important probe of the nature of
DM, and an important test of the favored cosmological model. 

Direct observation of dark substructure is by definition limited by its dark nature. 
As an effect that is directly sensitive to gravity regardless of the electromagnetic 
radiation, strong gravitational lensing has been shown to be a unique and powerful 
tool in probing the mass distribution in galaxy and cluster scales 
\citep[e.g.,][]{SLACSI, SLACSIII, SLACSII, SLACSIV, SLACSV, SLACSVI, SLACSVII, 
SLACSVIII, SLACSX, Bolton12a, Johnson14, Oguri2014, Sharon14, SLACSXII, Wang15}. 
In principle, dark substructures within lensing galaxies can induce 
observationally perceptible signals and therefore be indirectly detected. 
In particular, perturbations by dark substructures on the lensing magnification, 
which is sensitive to the second derivatives of the lensing potential, 
can occasionally become very strong. As first discussed by \citet{Mao98}, 
such perturbations on the magnification can be used as an explanation of the 
flux-ratio anomalies, which is the failure of smooth lens models in recovering 
the observed flux ratios of some lens systems 
\citep[e.g.,][]{Bradavc02, Dalal02, MacLeod13}. Note that similar effects can 
also be caused by low-mass structures along the line of sight 
\citep{Mao04, Metcalf05, Xu09, Xu12, Xu15, Inoue16a}. 
Astrometric perturbations by dark substructures, on the other hand, 
are usually too small for any substructure inference. 
If the surface brightness of the source further varies in time, 
then the time delays between multiple lensed images can be used as 
an extra test for the presence of substructures \citep[e.g.,][]{Keeton09}. 

Nevertheless, for extended static sources, the direct observable is the 
surface-brightness perturbation, a combination of both astrometric and magnification 
perturbations. By modeling such surface-brightness perturbations as a result of 
potential perturbations from dark substructures, recent
studies have found strong evidence for the presence of 
dark substructures in several lensing systems 
\citep[e.g.,][]{Vegetti10, Fadely12, Vegetti12, Nierenberg14, Hezaveh16, Inoue16b}. 
However, a large sample of such detections is needed for 
a statistical study of dark substructure. 

Motivated by the success of the Sloan Lens ACS Survey \citep[SLACS,][]{SLACSI}, 
the Baryon Oscillation Spectroscopic Survey (BOSS) Emission Line Lens Survey 
\citep[BELLS,][]{Brownstein2012}, and the SLACS for the Masses Survey 
\citep[S4TM,][]{SLACSXII} in uncovering significant new strong-lens samples, 
we initiated the BELLS for the GALaxy-Ly$\alpha$ EmitteR sYstems Survey 
(the \Survey{} Survey hereafter) for dark-substructure detections in galaxy-scale 
gravitational lenses with high-redshift Ly$\alpha$ emitters (LAEs) as the lensed 
sources. The lens candidates of the \Survey{} 
Survey are spectroscopically selected from the final data release, Data Release 12 
(DR12), of the BOSS \citep{Dawson13} of 
the Sloan Digital Sky Survey III \citep[SDSS-III,][]{SDSSIII}, using selection
techniques similar to those employed in the SLACS, BELLS, and S4TM surveys. 
A Hubble Space Telescope (\textsl{HST}) follow-up imaging
program has been approved and the observations are currently underway
(\textsl{HST} Cycle 23, GO Program \# 14189, PI: A. S. Bolton). The candidate systems
are massive galaxies at redshifts of approximately 0.5, whose BOSS spectra
show Ly$\alpha$ emission from more distant LAEs at redshifts of 2--3.

Extensive studies have demonstrated that the lower-mass limit of substructures 
that can produce observable lensing signals is related
to the characteristic size of the lensed source, with smaller sources being
sensitive to perturbations by smaller mass substructures 
\citep[e.g.,][]{Schneider92, Koopmans00, Moustakas03, Kochanek04, 
Mortonson05, Amara06}.
This is intuitively understandable by noting that smaller lensing masses have 
smaller associated lensing deflection angles, and lensed sources with smaller-scale 
structures subtending smaller angles will be correspondingly sensitive to the effects
of these smaller lensing masses. 
Furthermore, as discussed by \citet{Moustakas03}, the ``cut-off'' mass can be estimated 
as $m_c \propto \ell_s^{3/2}$, where $\ell_s$ is the characteristic
length scale of the lensed source. 
When observed at optical wavelengths,
high-redshift LAEs will show structure on small scales corresponding to
the far-UV emission from intense localized regions of star formation.
Lensed LAEs can hence push the detection limit of dark substructure 
down by roughly an order of magnitude in terms of its mass when comparing to 
the currently existing galaxy--galaxy lens samples in which the typical source size 
is a few times larger. 
Importantly, this lower-mass detection threshold can also raise the
number of expected substructure detections, since lower-mass dark
substructures are predicted to be significantly more abundant
than higher mass substructures 
\citep[e.g.,][]{Bullock01, DeLucia04, Gao04, vandenBosch05, Zentner05, Giocoli08}.
Additionally, the capability of detecting low-mass 
dark substructures is crucial for constraining the nature of DM because the 
differences in substructure properties predicted by CDM and other alternatives 
such as warm and self-interacting DM become significantly larger for lower 
substructure masses \citep[e.g.,][]{Colin00, Spergel00, AvilaReese01, Springel08, 
Lovell12, Lovell14, Bose16}.

The sample of lensed LAEs to be delivered by
the \Survey{} Survey will also be an invaluable resource 
for the study of high-redshift LAEs themselves.
LAEs are believed to be young, low-mass, extremely star-forming galaxies that 
serve as the progenitors of modern massive galaxies. They hold clues to the 
formation and evolution of galaxies at the time when the universe was still young. 
They can be used as tracers of the large-scale structure to constrain cosmological 
parameters \citep[e.g.,][]{Hill08}, and as a probe of the high-redshift 
intergalactic medium even across the reionization epoch 
\citep[e.g.,][]{Miralda-Escude98, Malhotra04, Zheng10, Zheng11}. 
The magnification effect of 
gravitational lensing in this new sample provides a ``cosmic telescope'' for these
objects, allowing us to probe their detailed structure below 
100 pc scales, which are otherwise beyond current observational capabilities. 

This paper describes the candidate selection strategy and 
presents the catalog of the 21 highest-quality galaxy--LAE lens candidates systems
being observed by \textsl{HST}, along with 
their properties as measured by the BOSS survey. Subsequent papers in
this series will present the results of the \textsl{HST} program.

\begin{figure}[htbp]
	\includegraphics[width=0.49\textwidth]{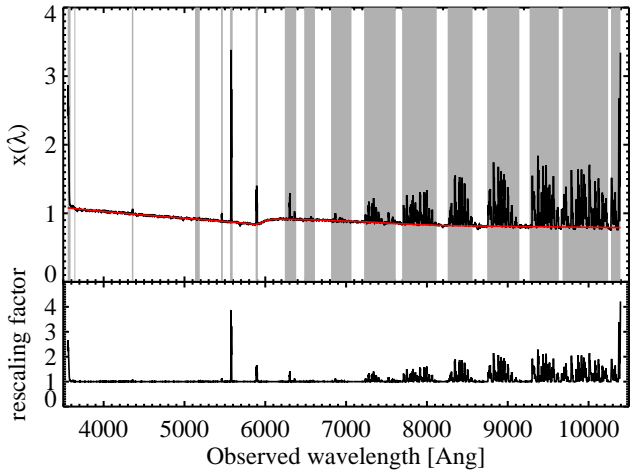}
\caption{\label{fig:renorm}
Top: error-weighted sky residual RMS spectrum (in black) and the B-spline 
fit (in red); Bottom: derived factor for error rescaling. 
The gray-shaded regions in the top panel indicate the masked wavelength ranges 
during the B-spline fit. 
}
\end{figure}

The outline of this paper is as follows. Section 2 describes in detail the 
lens candidate selection technique adopted for the \Survey{} Survey. 
Spectroscopic and photometric properties of the 21 highest-quality galaxy--LAE lens 
candidates are presented in Section 3. Strong-lensing probability is discussed 
in Section 4. Discussions on the unique studies enabled by this galaxy--LAE lens 
sample are given in Section 5. The final conclusion is in Section 6.

\section{Lens Candidate Identification}

The selection technique for the \Survey{} Survey is based on the
technique employed in the previous SLACS, BELLS, and the S4TM surveys, but with 
substantial modifications to specifically select lensed high-redshift LAEs. 
The basic principle of the selection technique is to identify a ``rogue''
emission line in the spectrum of a BOSS survey target galaxy, which can be an 
indication of a lensing configuration with two objects along 
the same line of sight.
A detailed description of the selection technique 
adopted for the \Survey{} Survey is presented in this section. This technique is
implemented by code that is released publicly with the publication of this paper.

\subsection{Flux-error Rescaling}

To ensure an accurate quantification of the detection significance of candidate 
rogue emission lines, the first step is to obtain an accurate estimate 
of the spectroscopic flux errors across the observed spectra. 
The BOSS spectroscopic pipeline reports noise vectors that are reasonably
accurate at most wavelengths, but underestimate the effects of
systematic sky-subtraction residuals at wavelengths with strong airglow lines.
Therefore, we recalibrate the reported noise vectors empirically using
the sky-subtracted data for the $\sim 100$ blank-sky fibers on every 
BOSS spectroscopic plate.

We define the parent sample for our search by selecting all plates
with ``good'' quality (as indicated by the \texttt{PLATEQUALITY} flag in SDSS 
terminology). For each of these plates, we select all the sky-model-subtracted
blank-sky spectra. We reject from our parent sample any plates that have a
root mean square (RMS) error-scaled residual flux across all sky spectra
of 1.4 or greater, taking this as an indication of unusually poor sky
subtraction. This RMS statistic would in principle be 1 for all spectra
if all noise were Gaussian, all errors were correctly estimated, and all sky
foregrounds were perfectly subtracted.

We then compute the RMS of the error-weighted blank-sky residuals across all
retained plates at each wavelength. The result is a spectrum of empirical values
$x(\lambda)$ (the black line in the top panel of Figure~\ref{fig:renorm}) 
indicating the factor by which errors are underestimated as a function of
wavelength. The easiest procedure for applying this spectrum would be to
simply inflate all reported errors by this wavelength-dependent factor $x$.
However, in some spectral ranges that are relatively free of sky emission lines,
the RMS error-weighted residual flux is less than unity as a result of
spectral rebinning effects. Since we wish to be conservative in our estimates
of detection significance and not \textit{deflate} any error estimates, we
fit a smooth B-spline model to the baseline trend of $x(\lambda)$, masking
the spectral ranges that are significantly affected by airglow lines. 
The red line in the top panel of Figure~\ref{fig:renorm} represents the B-spline fit, 
and the gray-shaded regions visualize the masked spectral ranges. 
We divide $x(\lambda)$ by this baseline fit to obtain the factor by which 
we rescale all our error estimates in subsequent analysis steps. 
The bottom panel in Figure~\ref{fig:renorm} shows the resulting rescaling factor 
as a function of wavelength. 
Note that the rescaling factors are forced to be at least unity.


\begin{table*}[htbp]
\begin{center}
\caption{\label{tb:targets} Selected properties of the 21 galaxy--LAE lens candidate systems.}
\begin{tabular}{l c c c c c c c}
\hline \hline
Target & Plate--MJD--Fiber & $z_{L}$ & $z_{s}$ & R.A. & Decl. & $m_i^{a}$ & Ly$\alpha$ Flux$^{b}$ \\
\hline
SDSS\,J002927.38$+$254401.7 & 6281-56295-811 & 0.5869 & 2.4504 & 00 29 27.3883 & $+$25 44 01.7862 & 19.22 & 32.88 \\
\hline
SDSS\,J005409.97$+$294450.8 & 6254-56268- 35 & 0.4488 & 2.7176 & 00 54 09.9754 & $+$29 44 50.8882 & 19.53 & 26.48 \\
\hline
SDSS\,J011300.57$+$025046.2 & 4315-55503-703 & 0.6230 & 2.6088 & 01 13 00.5750 & $+$02 50 46.2048 & 19.45 & 29.66 \\
\hline
SDSS\,J020121.39$+$322829.6$^{c}$ & 6605-56565-722 & 0.3957 & 2.8209 & 02 01 21.3954 & $+$32 28 29.6649 & 18.32 & 26.43 \\
\hline
SDSS\,J023740.63$-$064112.9 & 4399-55811-149 & 0.4859 & 2.2491 & 02 37 40.6393 & $-$06 41 12.9900 & 19.22 & 32.27 \\
\hline
SDSS\,J074249.68$+$334148.9 & 4440-55539- 41 & 0.4936 & 2.3633 & 07 42 49.6857 & $+$33 41 48.9890 & 19.47 & 45.40 \\
\hline
SDSS\,J075523.52$+$344539.5$^{c}$ & 3754-55488-639 & 0.7224 & 2.6347 & 07 55 23.5254 & $+$34 45 39.5920 & 20.54 & 11.29 \\
\hline
SDSS\,J085621.59$+$201040.5 & 5175-55955-755 & 0.5074 & 2.2335 & 08 56 21.5918 & $+$20 10 40.5510 & 19.44 & 33.56 \\
\hline
SDSS\,J091807.86$+$451856.7 & 5813-56363-341 & 0.5238 & 2.3440 & 09 18 07.8662 & $+$45 18 56.7691 & 19.35 & 35.83 \\
\hline
SDSS\,J091859.21$+$510452.5$^{d}$ & 5730-56607-697 & 0.5811 & 2.4030 & 09 18 59.2126 & $+$51 04 52.5934 & 19.71 & 22.93 \\
\hline
SDSS\,J111027.11$+$280838.4 & 6435-56341-855 & 0.6073 & 2.3999 & 11 10 27.1106 & $+$28 08 38.4654 & 19.91 & 28.65 \\
\hline
SDSS\,J111040.42$+$364924.4 & 4622-55629-615 & 0.7330 & 2.5024 & 11 10 40.4260 & $+$36 49 24.3997 & 19.86 & 33.18 \\
\hline
SDSS\,J111634.55$+$091503.0 & 5369-56272-541 & 0.5501 & 2.4536 & 11 16 34.5593 & $+$09 15 03.0453 & 19.40 & 46.61 \\
\hline
SDSS\,J114154.71$+$221628.8 & 6422-56328-303 & 0.5858 & 2.7624 & 11 41 54.7119 & $+$22 16 28.8931 & 19.73 & 20.34 \\
\hline
SDSS\,J120159.02$+$474323.2$^{d}$ & 6672-56386-474 & 0.5628 & 2.1258 & 12 01 59.0259 & $+$47 43 23.1967 & 19.30 & 182.01 \\
\hline
SDSS\,J122656.45$+$545739.0 & 6833-56413-181 & 0.4980 & 2.7322 & 12 26 56.4587 & $+$54 57 39.0454 & 18.88 & 36.71 \\
\hline
SDSS\,J151641.22$+$495440.7 & 6727-56369-967 & 0.5479 & 2.8723 & 15 16 41.2207 & $+$49 54 40.7785 & 19.02 & 32.68 \\
\hline
SDSS\,J152926.41$+$401548.8 & 5165-56063-315 & 0.5308 & 2.7920 & 15 29 26.4148 & $+$40 15 48.8480 & 19.35 & 30.87 \\
\hline
SDSS\,J222825.76$+$120503.9 & 5048-56218-801 & 0.5305 & 2.8324 & 22 28 25.7666 & $+$12 05 03.9505 & 19.89 & 24.36 \\
\hline
SDSS\,J224505.93$+$004018.3 & 4205-55454-645 & 0.7021 & 2.5413 & 22 45 05.9326 & $+$00 40 18.3506 & 19.56 & 27.21 \\
\hline
SDSS\,J234248.68$-$012032.5 & 4356-55829-547 & 0.5270 & 2.2649 & 23 42 48.6841 & $-$01 20 32.5351 & 19.57 & 54.95 \\
\hline \hline
\end{tabular}
\end{center}
\smallskip
$a$: BOSS-measured \emph{i}-band apparent cmodel magnitude within the $1^{\arcsec}$ fiber. \\
$b$: Total apparent flux of the Ly$\alpha$ emission in units of $10^{-17}$\,erg\,cm$^{-2}$\,s$^{-1}$. \\
$c$: Two systems with strong evidence for lensing signals in their SDSS images. Please see Section~\ref{sect:notes} for details. \\
$d$: Two systems with probable evidence for lensing signals in their SDSS images. Please see Section~\ref{sect:notes} for details.
\end{table*}

\subsection{Emission-line Search}

After applying our flux-error rescaling, we search for rogue emission lines 
in the spectra of galaxies. The $\sim 1.5\times10^6$ optical galaxy spectra in 
the SDSS-III BOSS database, with resolution $R\approx 2000$ and wavelength
coverage from $3500$ to 10,400\,\AA\@, provide an unmatched sample
for this search. We select the subset from the BOSS final data release, DR12, 
that have confidently measured redshifts and classifications
as galaxies according to the automated analysis pipeline \citep{Bolton12b}. 
That results in 1,388,190 records. 
We subtract from each spectrum the pipeline's best-fit galaxy template
to obtain the galaxy-subtracted residual spectrum. 
We then perform an error-weighted matched-filter search within the galaxy-subtracted 
residual spectrum with a Gaussian kernel of 150\,km\,s$^{-1}$ dispersion, 
and retain detections (``hits'') of 6-$\sigma$ significance (signal-to-noise ratio, SNR) 
and above \citep{Bolton04}. We confine our search to the observed wavelength range
3600\AA~$< \lambda <$4800\,\AA\ (roughly $2 < z < 3$ for the Ly$\alpha$ emission)
so that low-redshift H$\alpha$, [O\textsc{iii}]~5007, and H$\beta$ cannot
be detected as interlopers masquerading as high-redshift Ly$\alpha$.
This procedure identifies 4982 hits in total. 

\subsection{False Positives Removal}

We apply several restrictive cuts to remove spurious detections. 
We first reject any significant overdensities of hits in both observed wavelength
(associated with airglow features) and BOSS target-galaxy
rest wavelength (associated with template-subtraction residuals). This cut removes 
almost $65\%$ (3212) of the total 4982 hits.  
To reject low-redshift [O\textsc{ii}]~3727 interlopers, we compute 
emission-line detection significance at wavelength positions
that would correspond to H$\alpha$, 
[O\textsc{iii}]~5007, and H$\beta$ if the primary ``hit'' were [O\textsc{ii}]~3727
instead of Ly$\alpha$. 
We drop any hit for which the quadrature sum of SNR values at these
three wavelength positions is greater than 2.5$\sigma$. This step further removes 845 hits. 
We then fit the spectrum of each remaining emission-line hit with a pixel-integrated 
skew-normal profile (Gaussian times error function) in order to quantify 
the line flux, line width, and skewness.
The spectra of all 290 hits with significance greater than 8$\sigma$ are
inspected visually, to reject detections that are associated
with either the subtraction of an incorrect redshift template or cross-talk
from a neighboring spectrum on the CCD detector.
The procedure yields a final parent sample of 187 candidate systems, 
which are further inspected by 
(1) examining their individual 15-minute sub-exposure spectra 
in order to ensure that the detection does not come from one 
sub-exposure alone, (2) examining the raw spectrograph CCD frames 
to ensure that the detections are not associated with cosmetic CCD 
defects, (3) examining the flat-fielding calibration 
vectors applied to the spectra at the wavelength of the detection, 
to ensure that no flat-fielding artifacts have been imprinted 
onto the spectra, and (4) examining their SDSS 
color images. None of the remaining systems are rejected by these final checks.

\subsection{Final Sample for \textsl{HST} Follow-up}

To make the best use of \textsl{HST} time and maximize the detection success rate, 
a highest-quality subsample was selected from the final parent sample. 
We first restrict the foreground lens to be a BOSS CMASS galaxy. 
The detailed magnitude and color cuts used for the classifications of 
CMASS and LOWZ galaxies can be found in \citep[][]{Eisenstein11, Shu12, Dawson13}. 
The subsample of candidates that we propose for
\textsl{HST} observation are selected to have a detection significance of 
15$\sigma$ or greater, an observed emission-line flux greater than 
$2\times 10^{-16}$\,erg\,cm$^{-2}$\,s$^{-1}$, and evidence for positive skewness 
(the classic Ly$\alpha$ ``blue edge, red tail'' profile)
of $\Delta \chi^2 > 4$ relative to the best-fit symmetric Gaussian
line profile. Two of these purely spectroscopically selected systems
also show indications of blue lensed features in the SDSS images. 
In addition, we also include two systems in the longer candidate list 
which did not meet these more stringent spectroscopic requirements 
but showed obvious signatures of blue lensed features in their SDSS image 
(See Section~\ref{sect:notes} for details). 
The list of the final 21 highest-quality galaxy--LAE candidates 
selected by these final cuts is presented in Table~\ref{tb:targets}.

\section{The \Survey{} Sample}

In this section, we present the BOSS-determined spectroscopic and photometric  
properties of the 21 galaxy--LAE lens candidate systems 
targeted for \textsl{HST} follow-up observations. 

\begin{figure*}[htbp]
\centering
\includegraphics[width=0.33\textwidth]{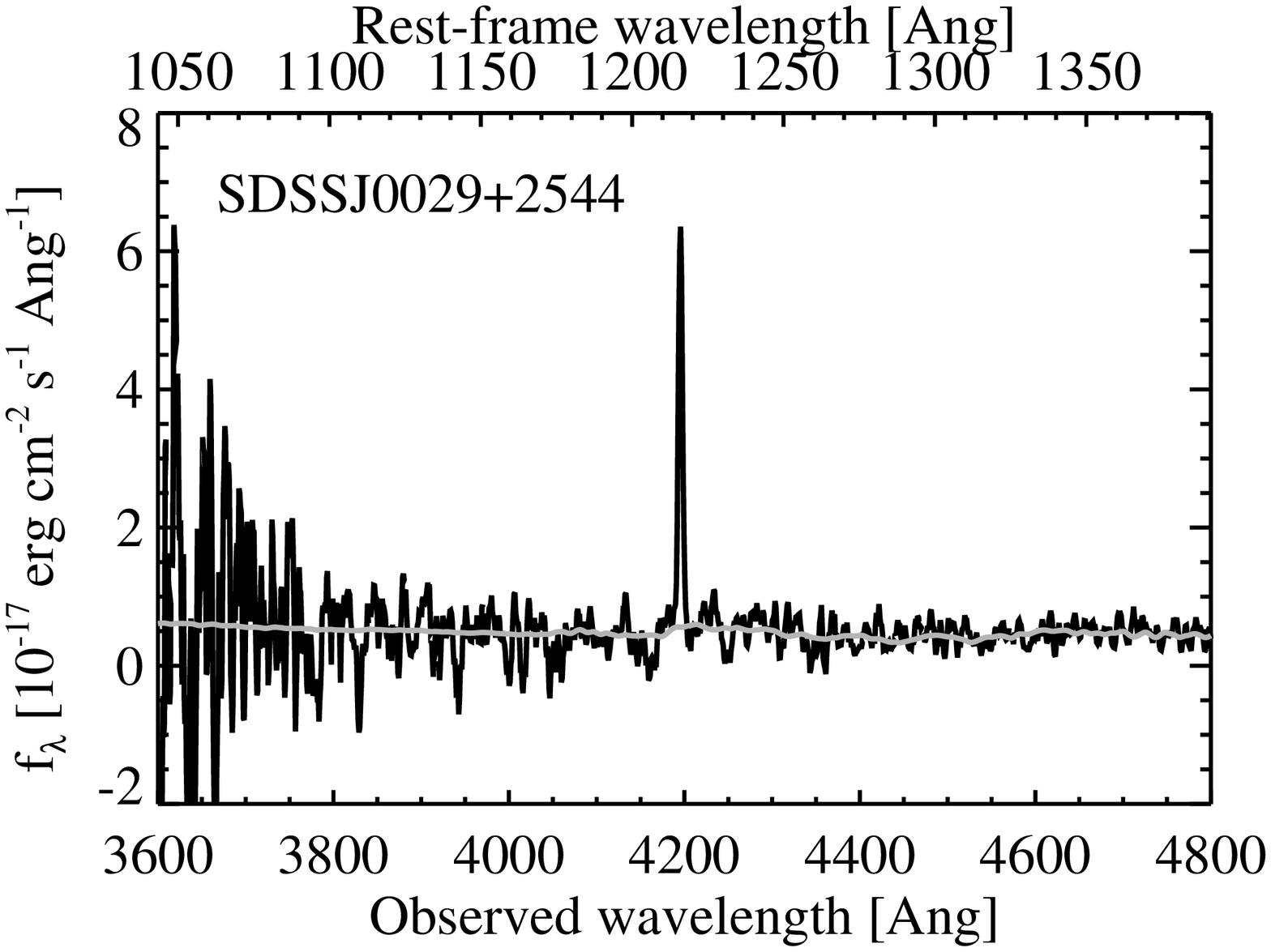}
\includegraphics[width=0.33\textwidth]{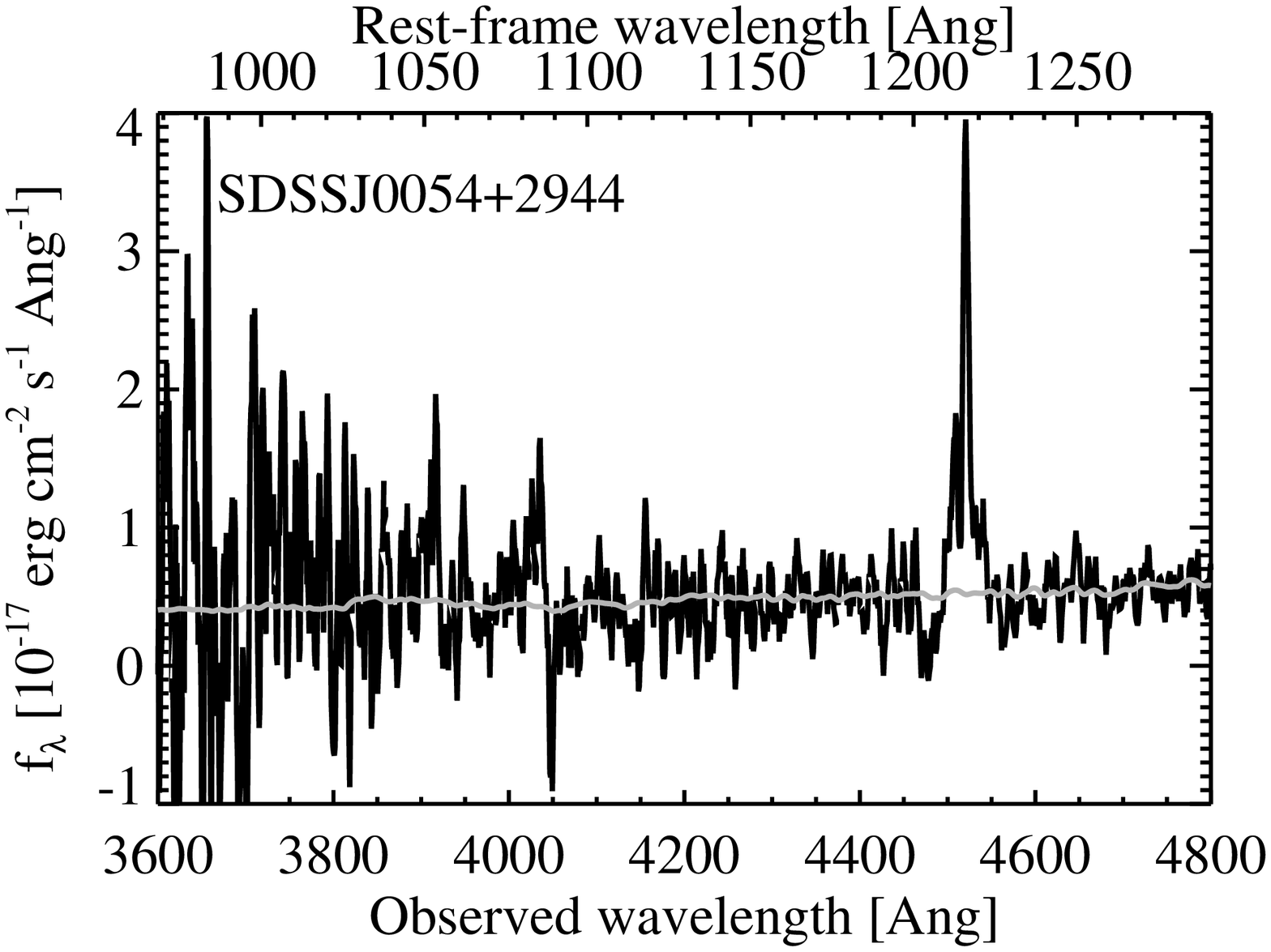}
\includegraphics[width=0.33\textwidth]{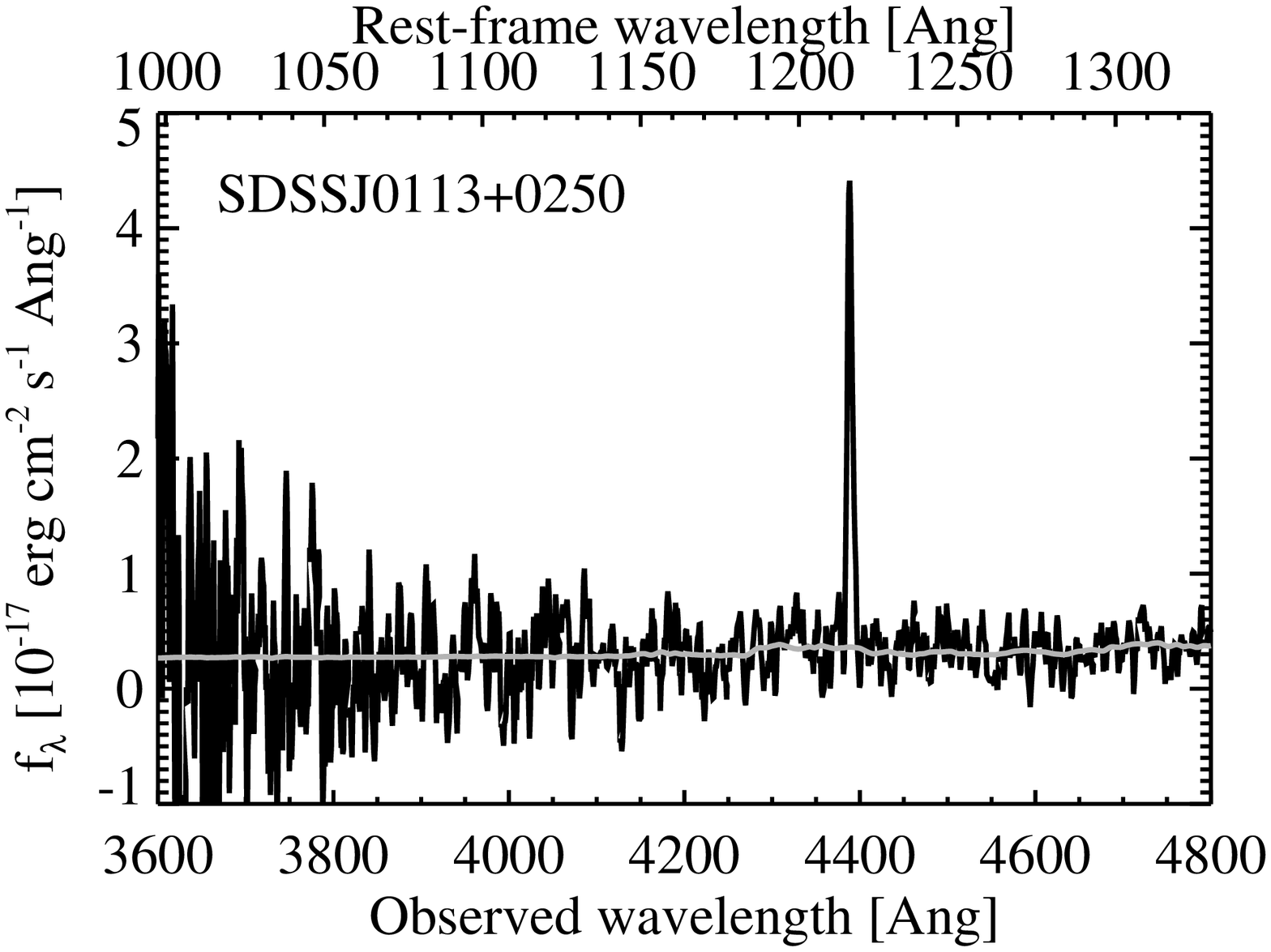}
\includegraphics[width=0.33\textwidth]{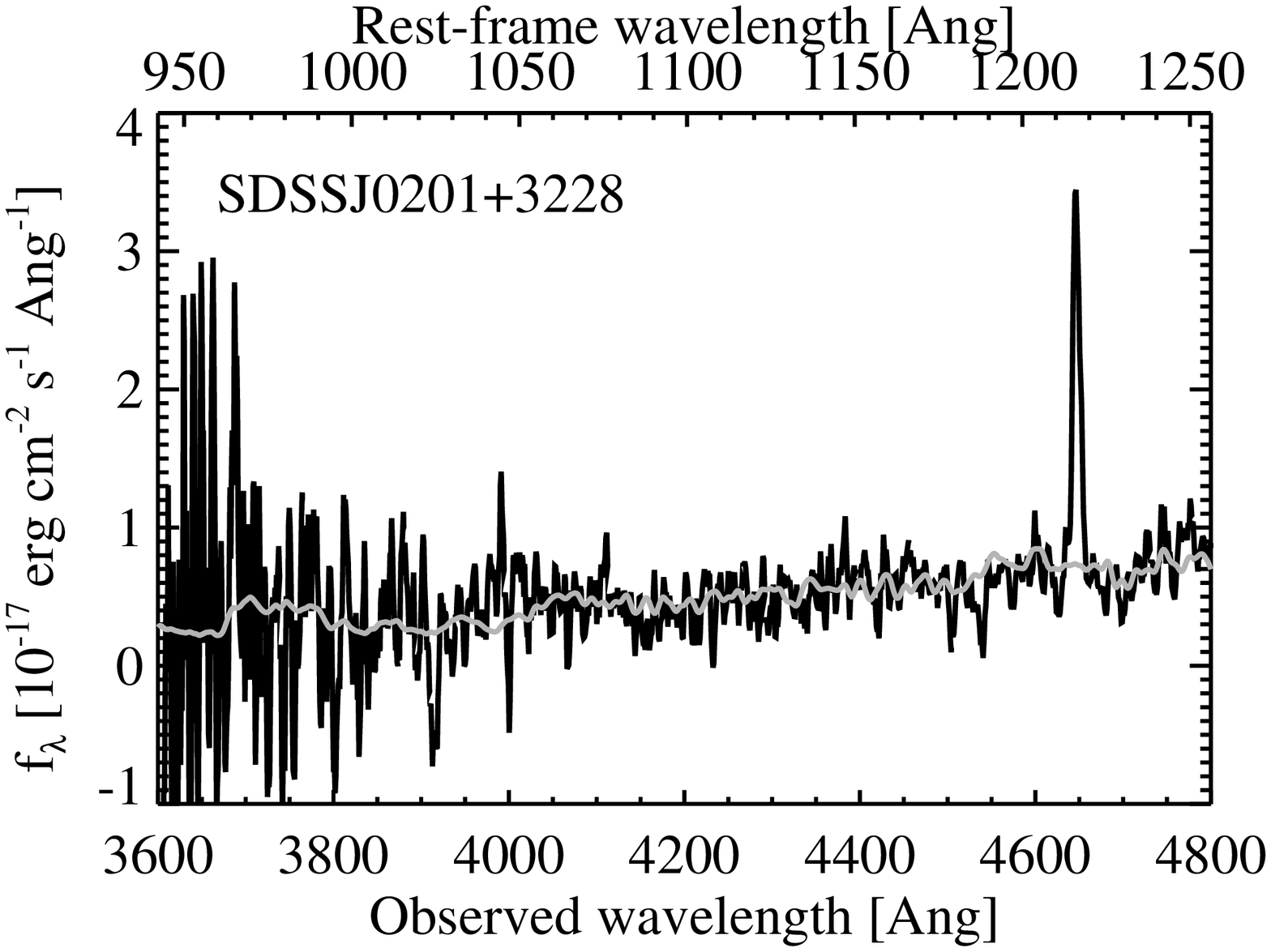}
\includegraphics[width=0.33\textwidth]{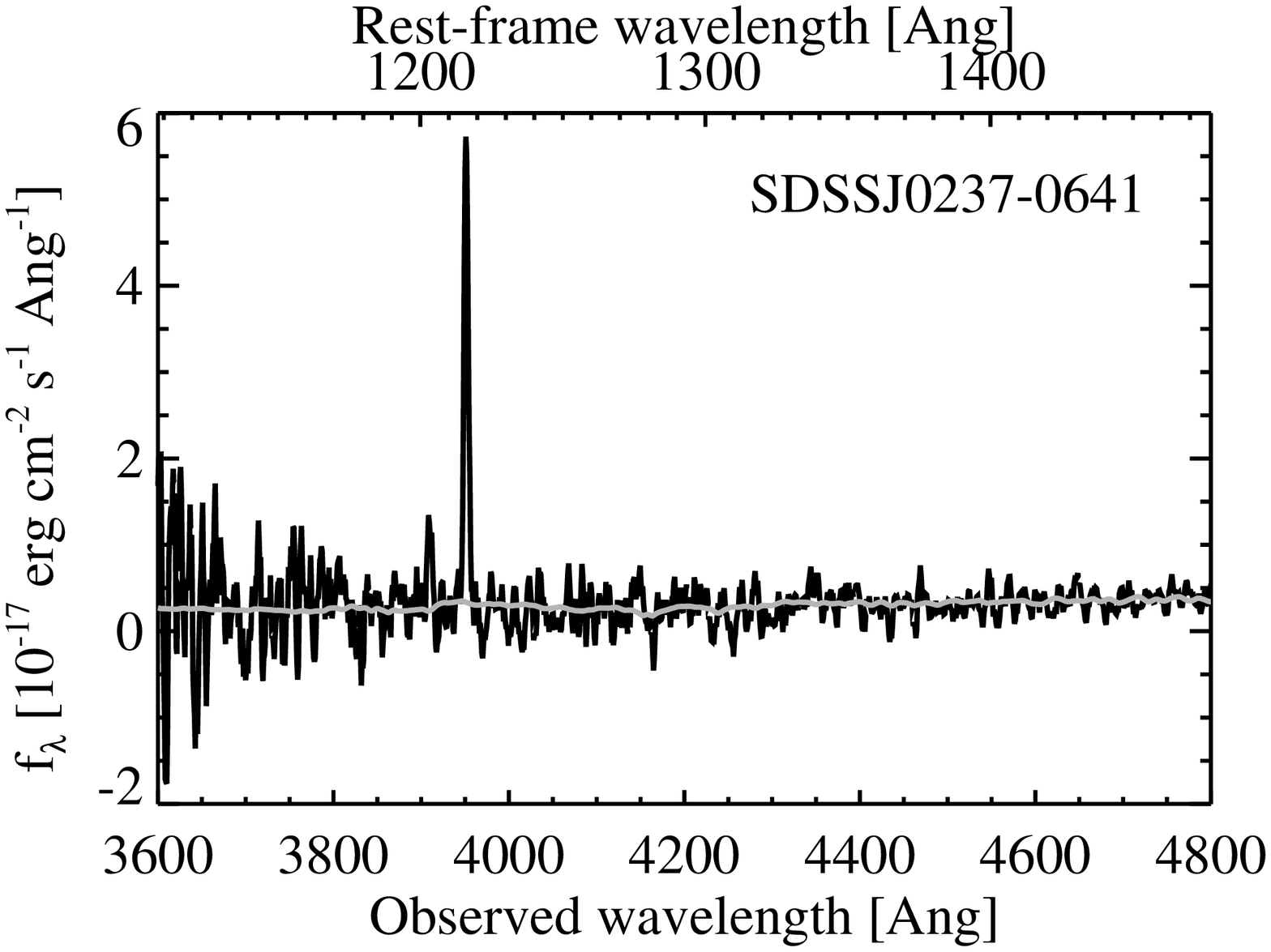}
\includegraphics[width=0.33\textwidth]{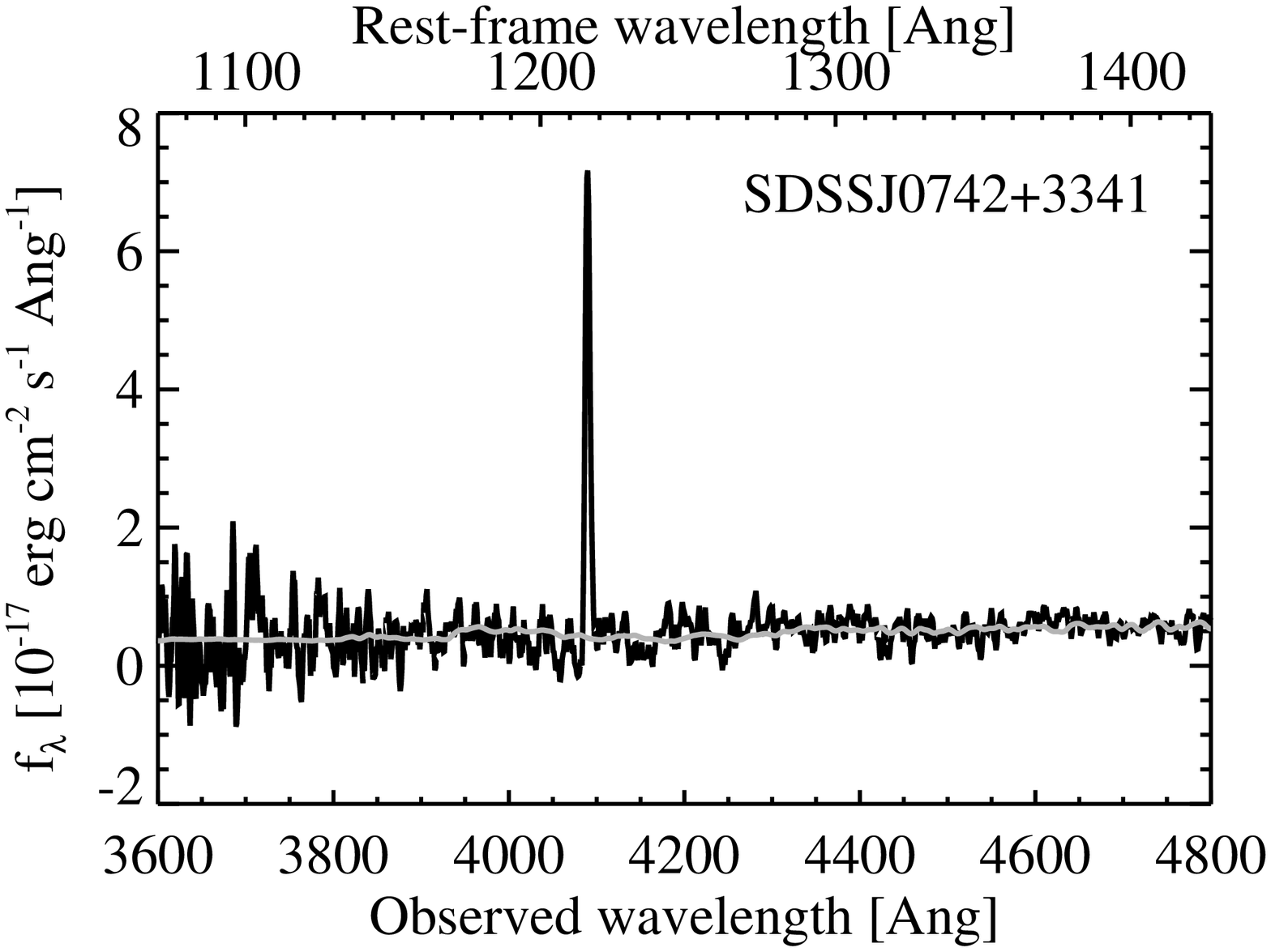}
\includegraphics[width=0.33\textwidth]{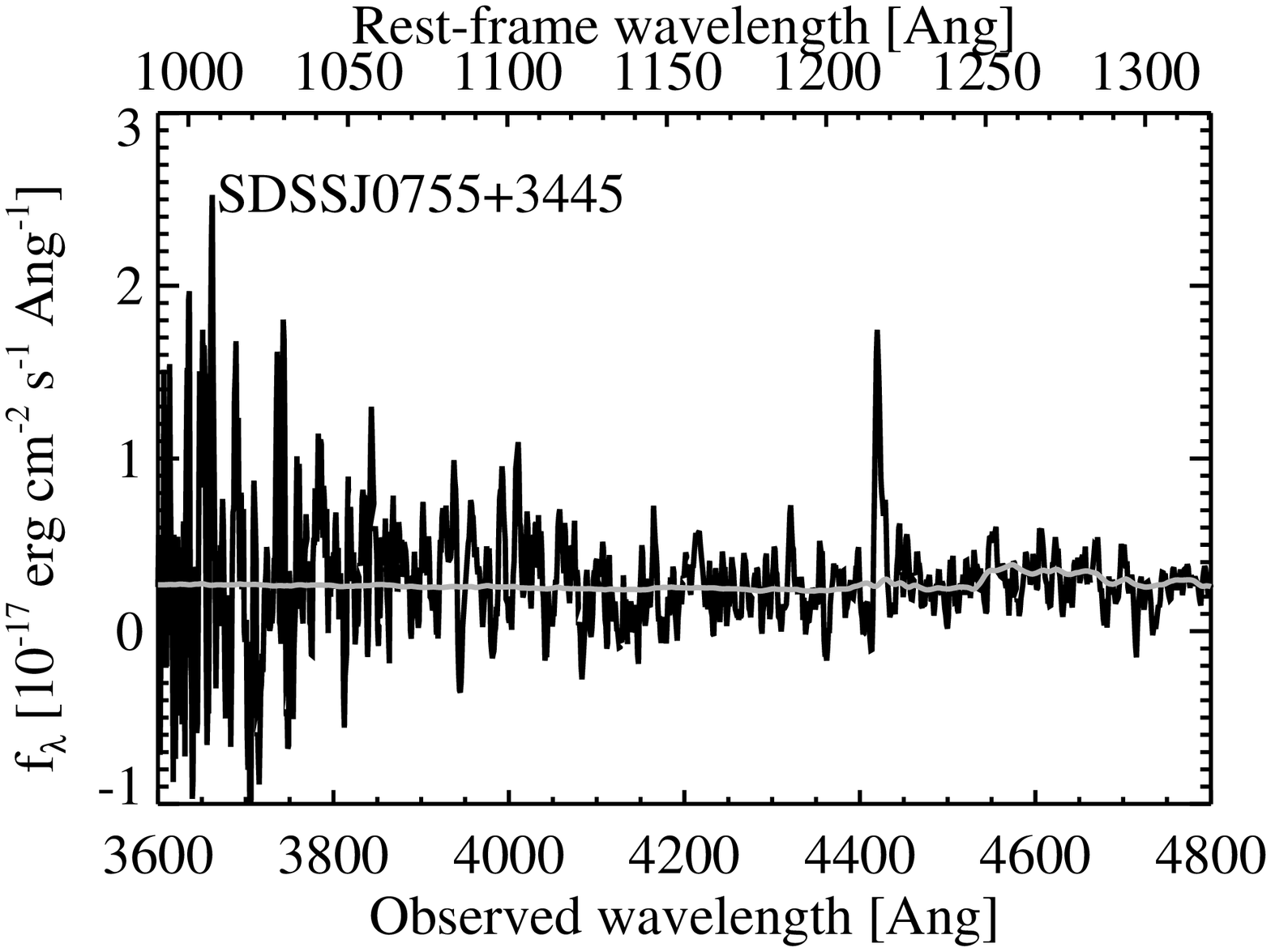}
\includegraphics[width=0.33\textwidth]{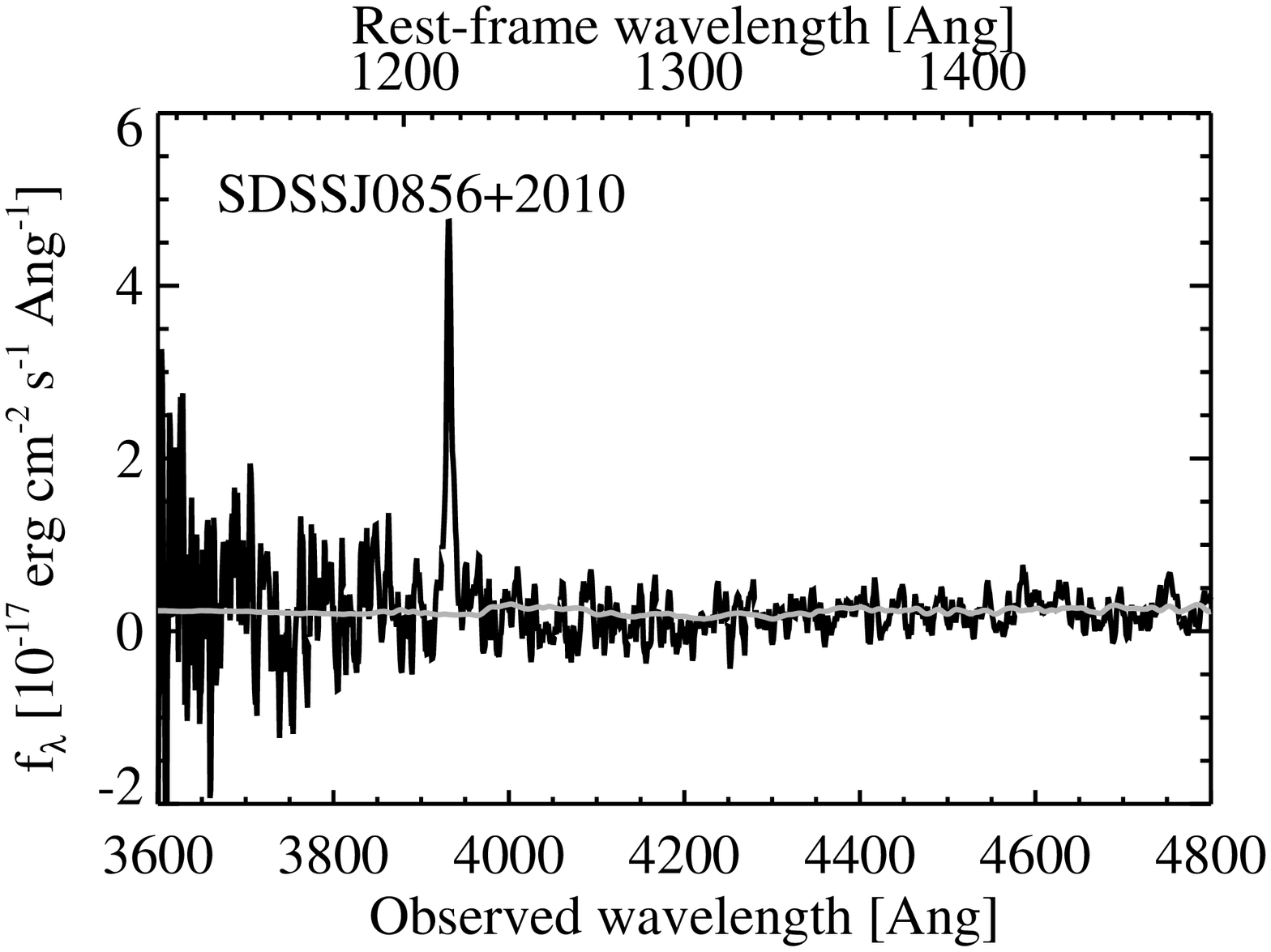}
\includegraphics[width=0.33\textwidth]{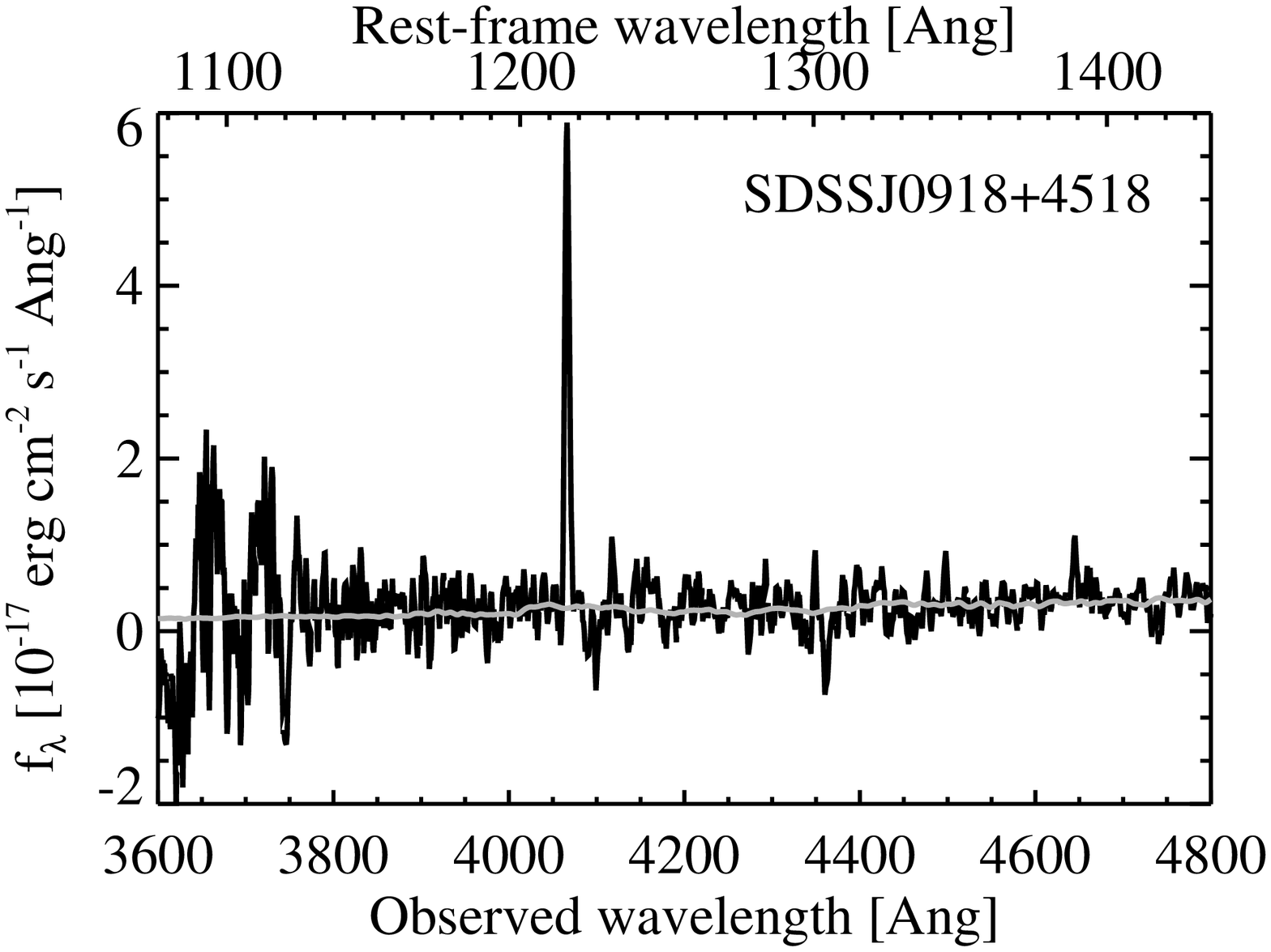}
\includegraphics[width=0.33\textwidth]{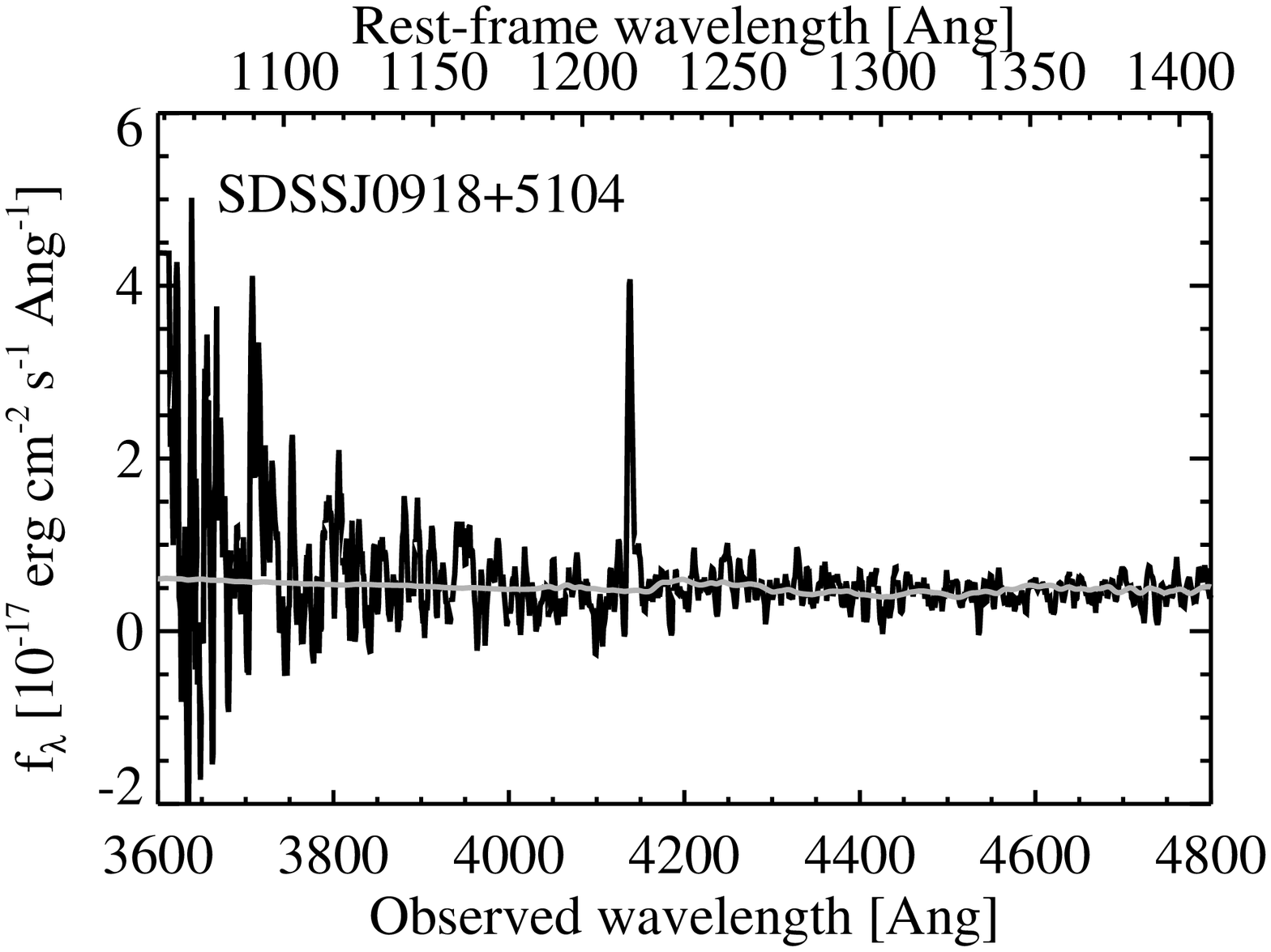}
\includegraphics[width=0.33\textwidth]{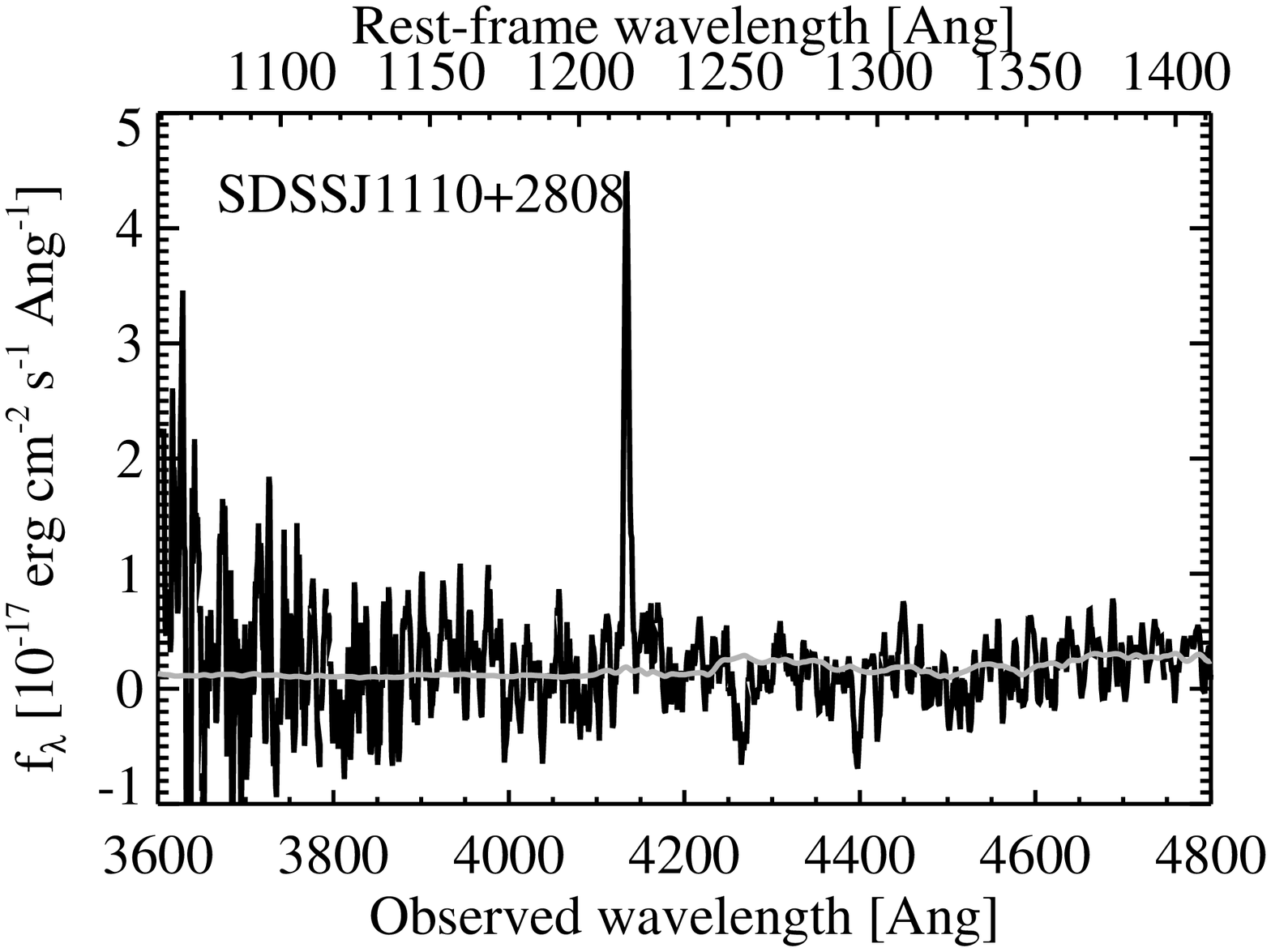}
\includegraphics[width=0.33\textwidth]{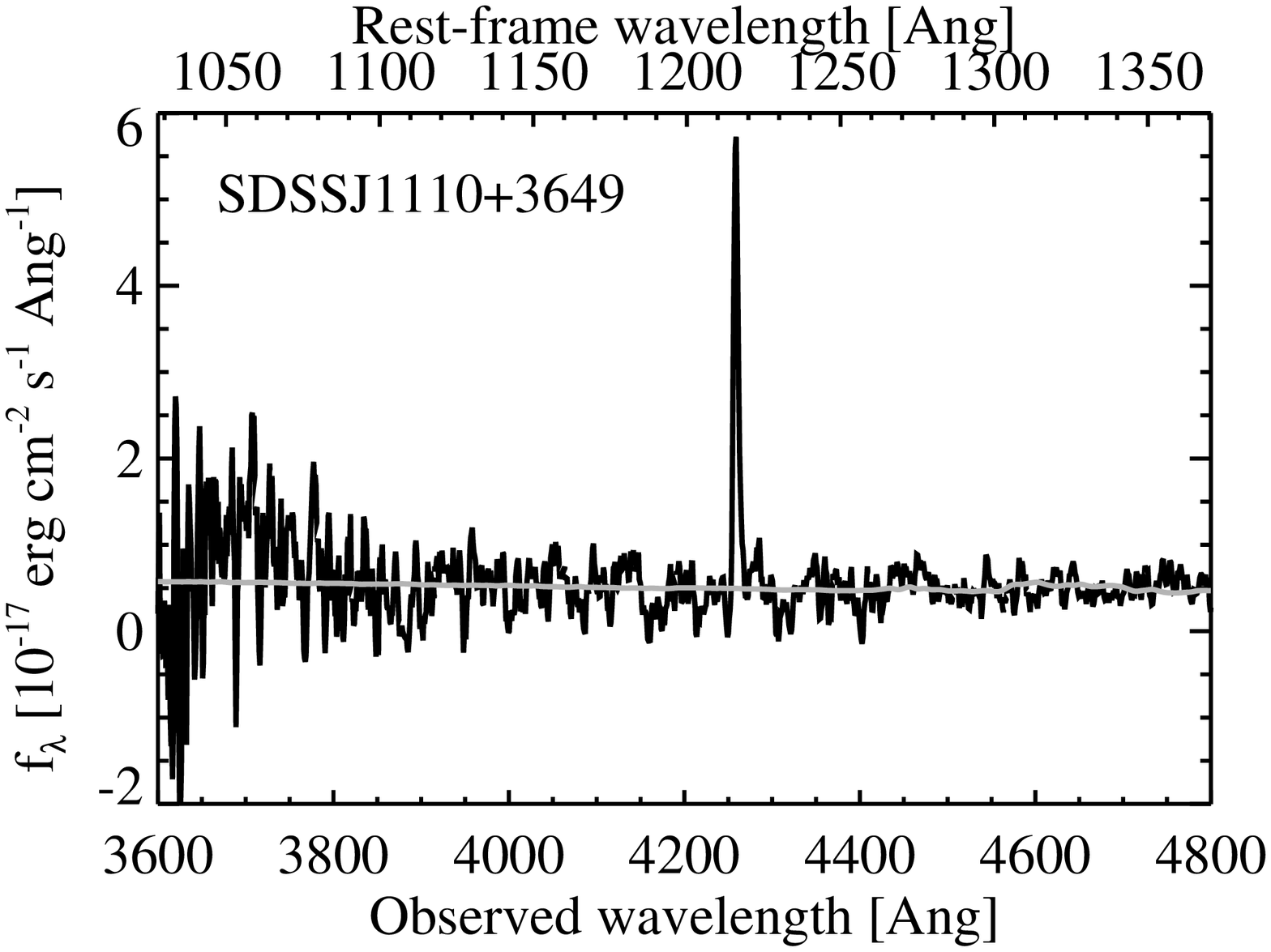}
\includegraphics[width=0.33\textwidth]{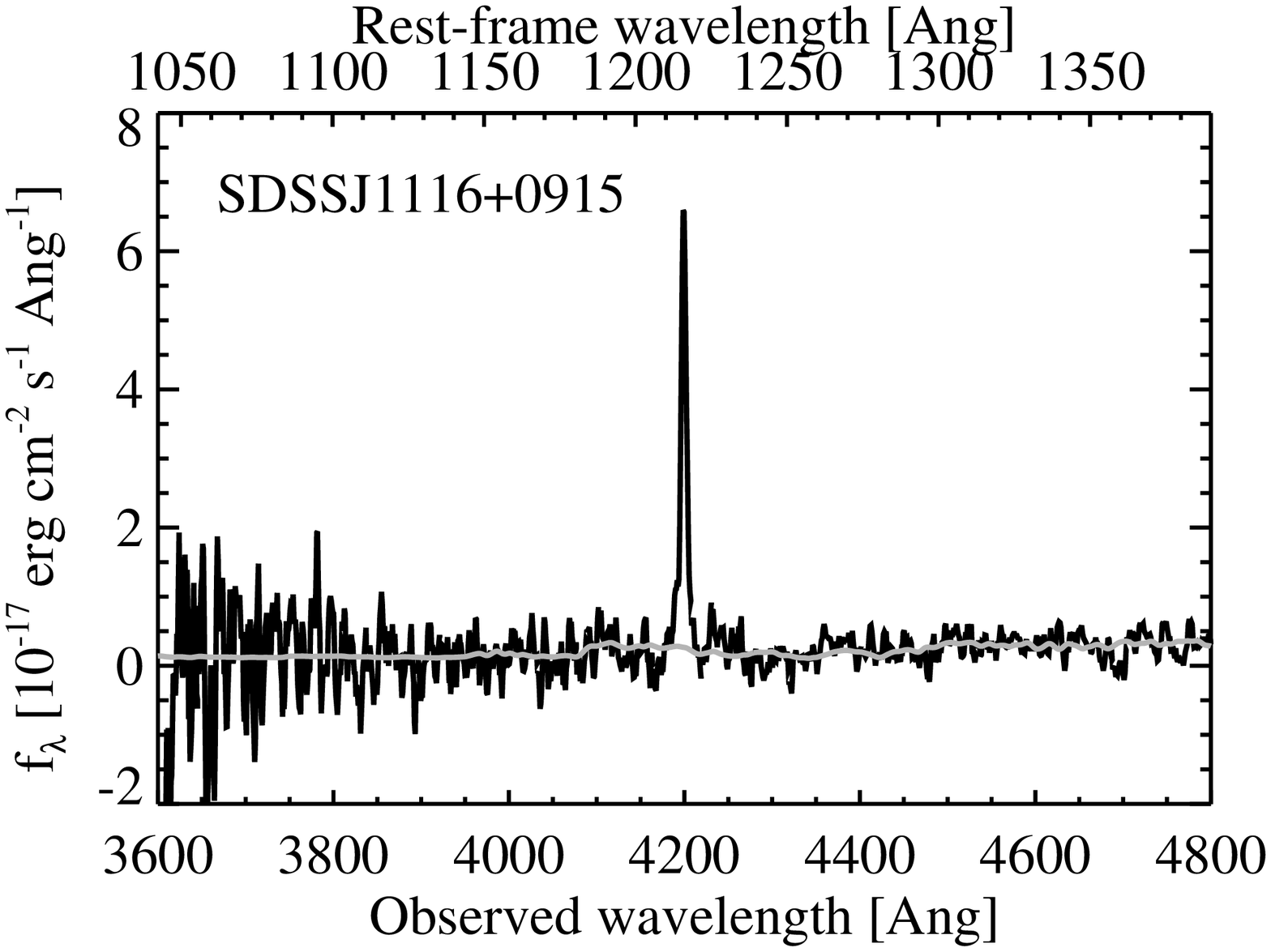}
\includegraphics[width=0.33\textwidth]{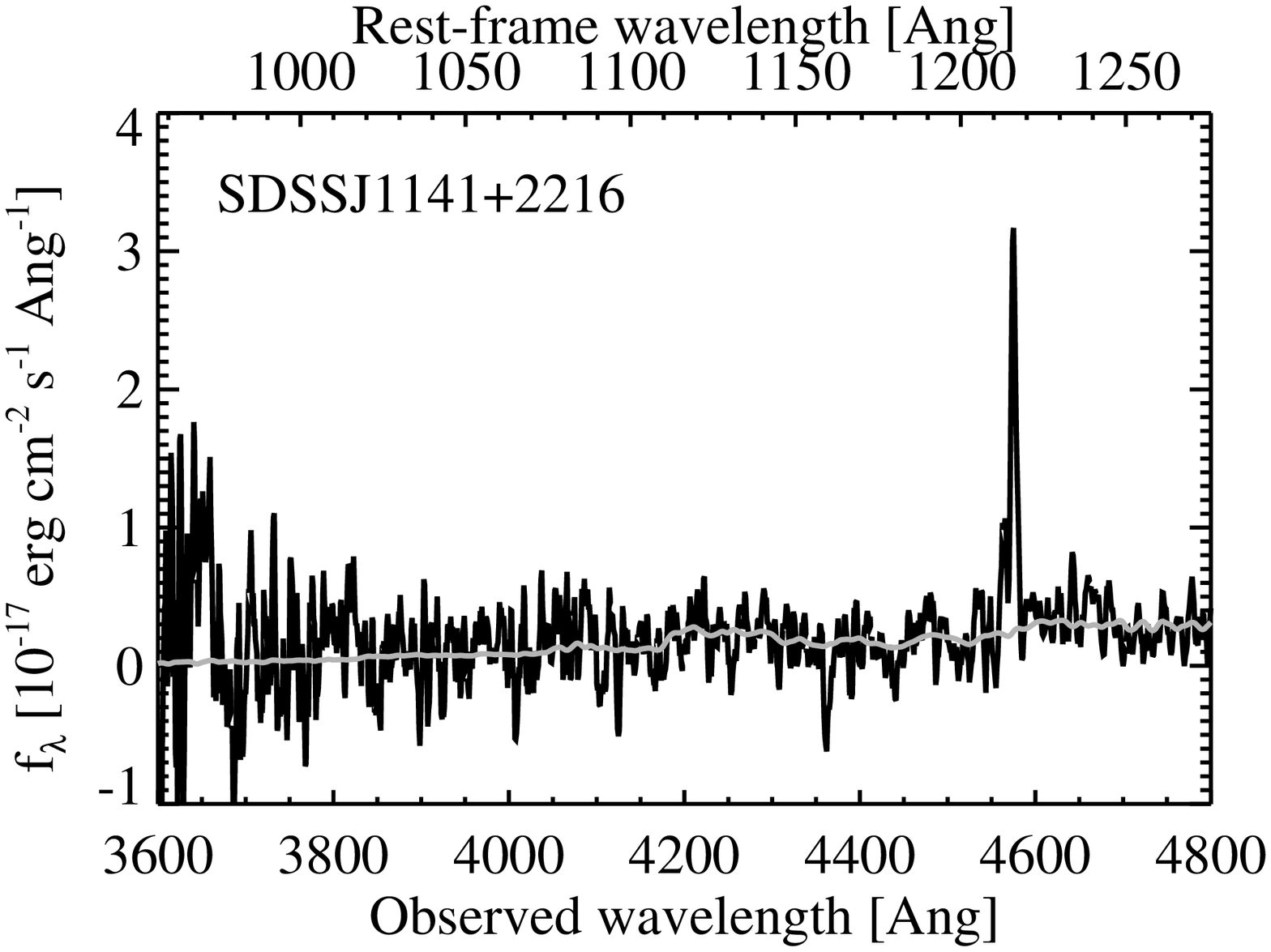}
\includegraphics[width=0.33\textwidth]{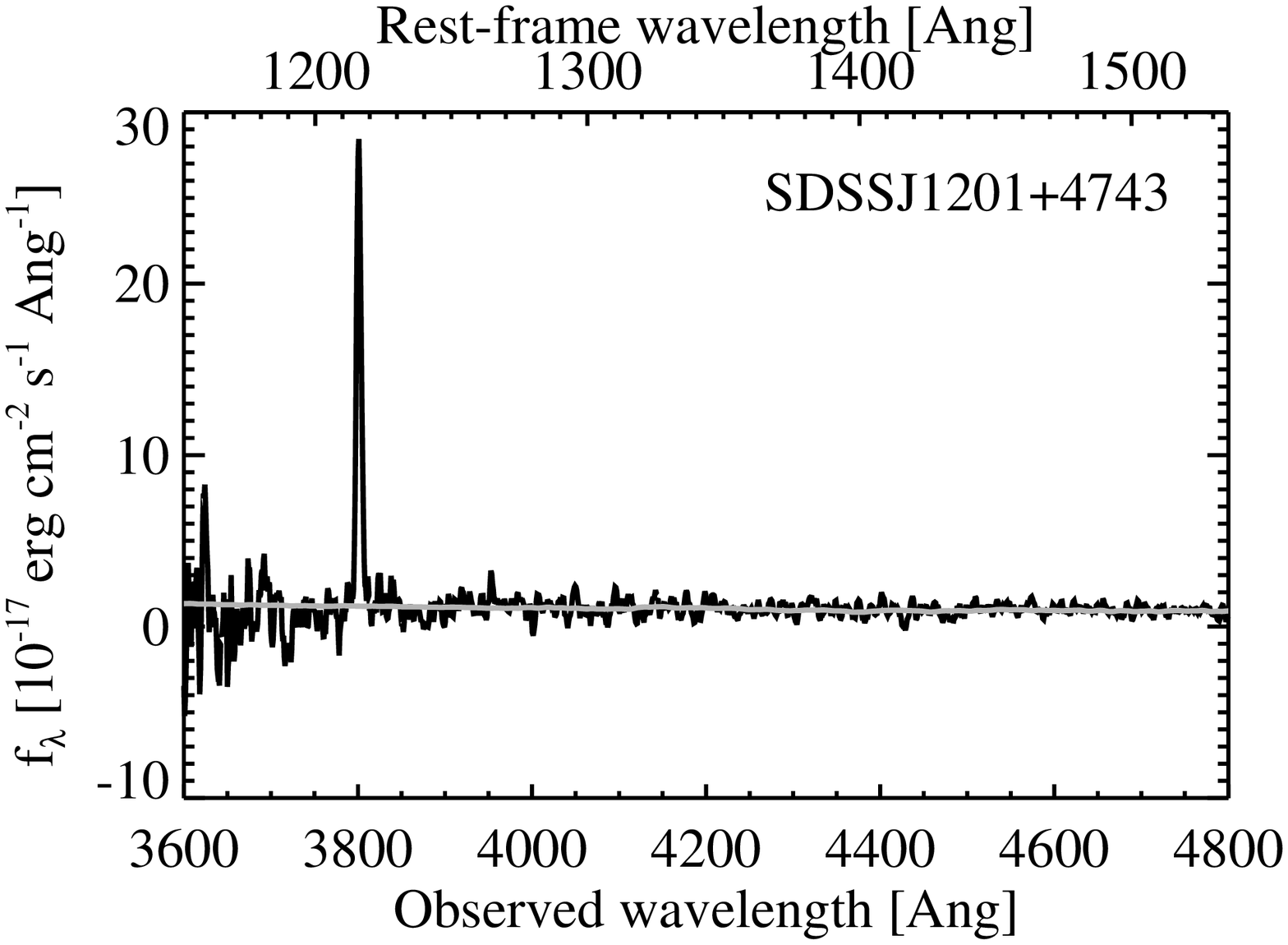}
\caption{{\label{fig:spectra1}} Smoothed BOSS spectra of the \Survey{} sample. Gray lines represent the best-fit continuum flux for the foreground galaxies.}
\end{figure*}
\addtocounter{figure}{-1}
\begin{figure*}[htbp]
\includegraphics[width=0.33\textwidth]{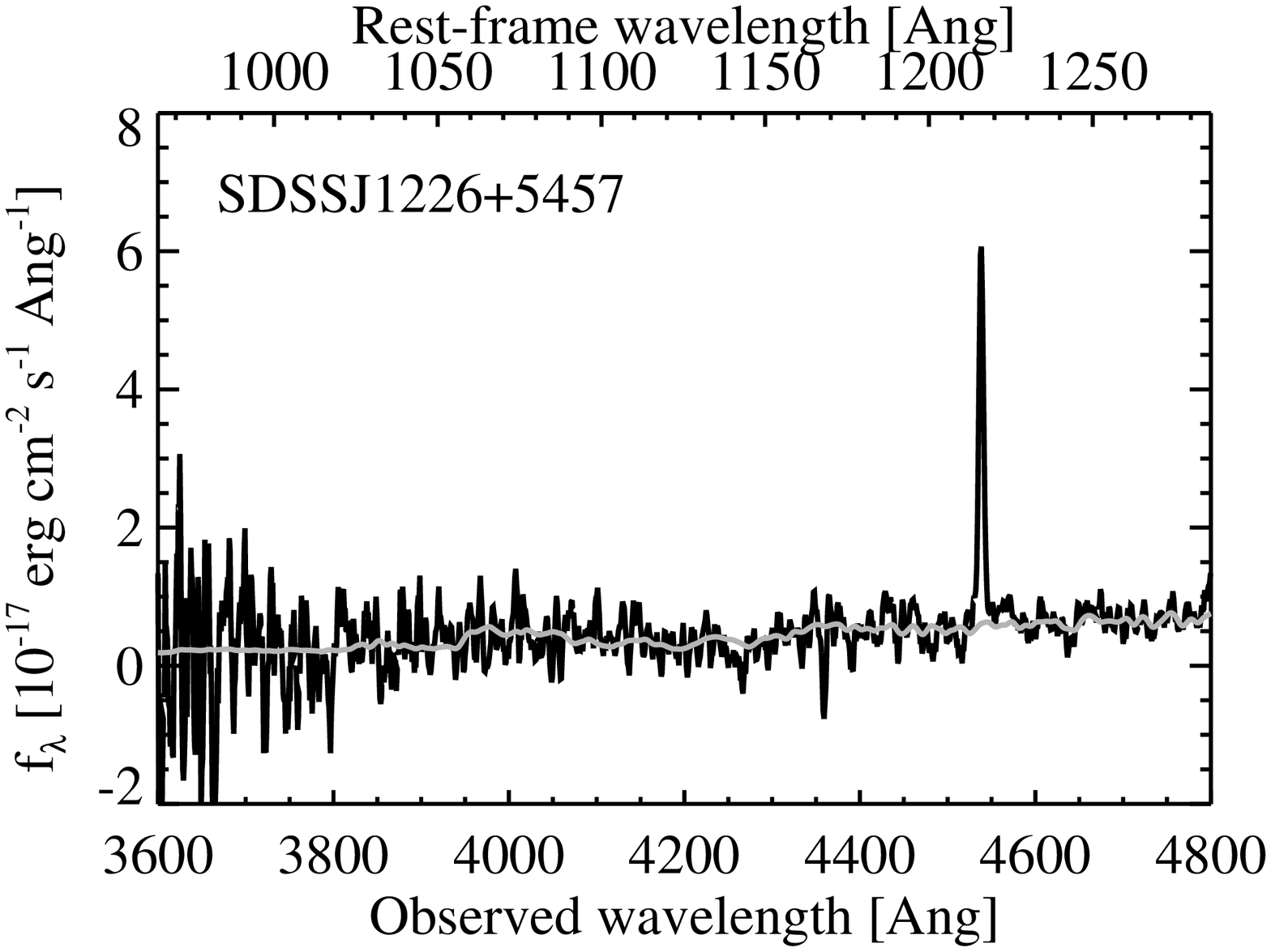}
\includegraphics[width=0.33\textwidth]{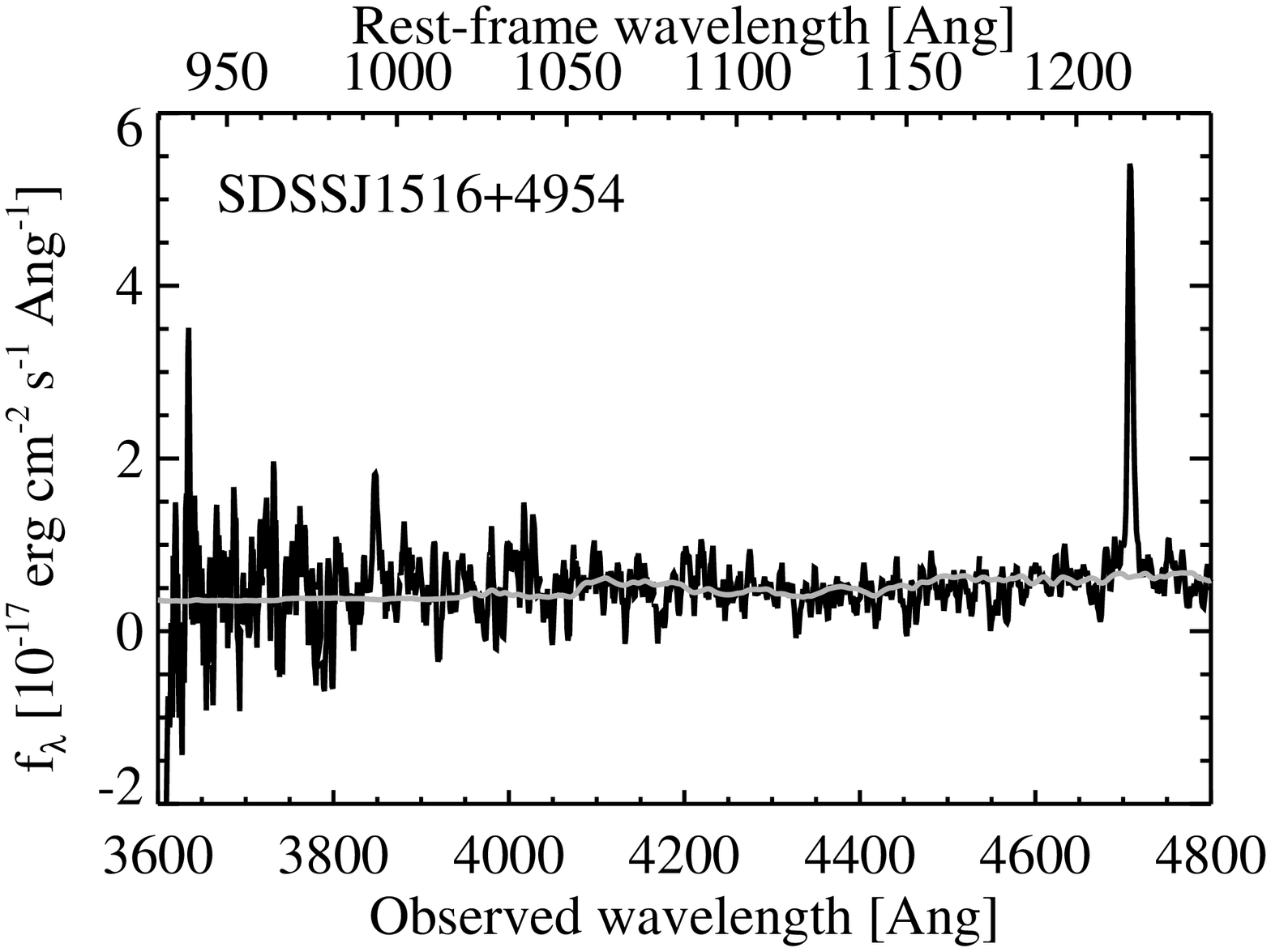}
\includegraphics[width=0.33\textwidth]{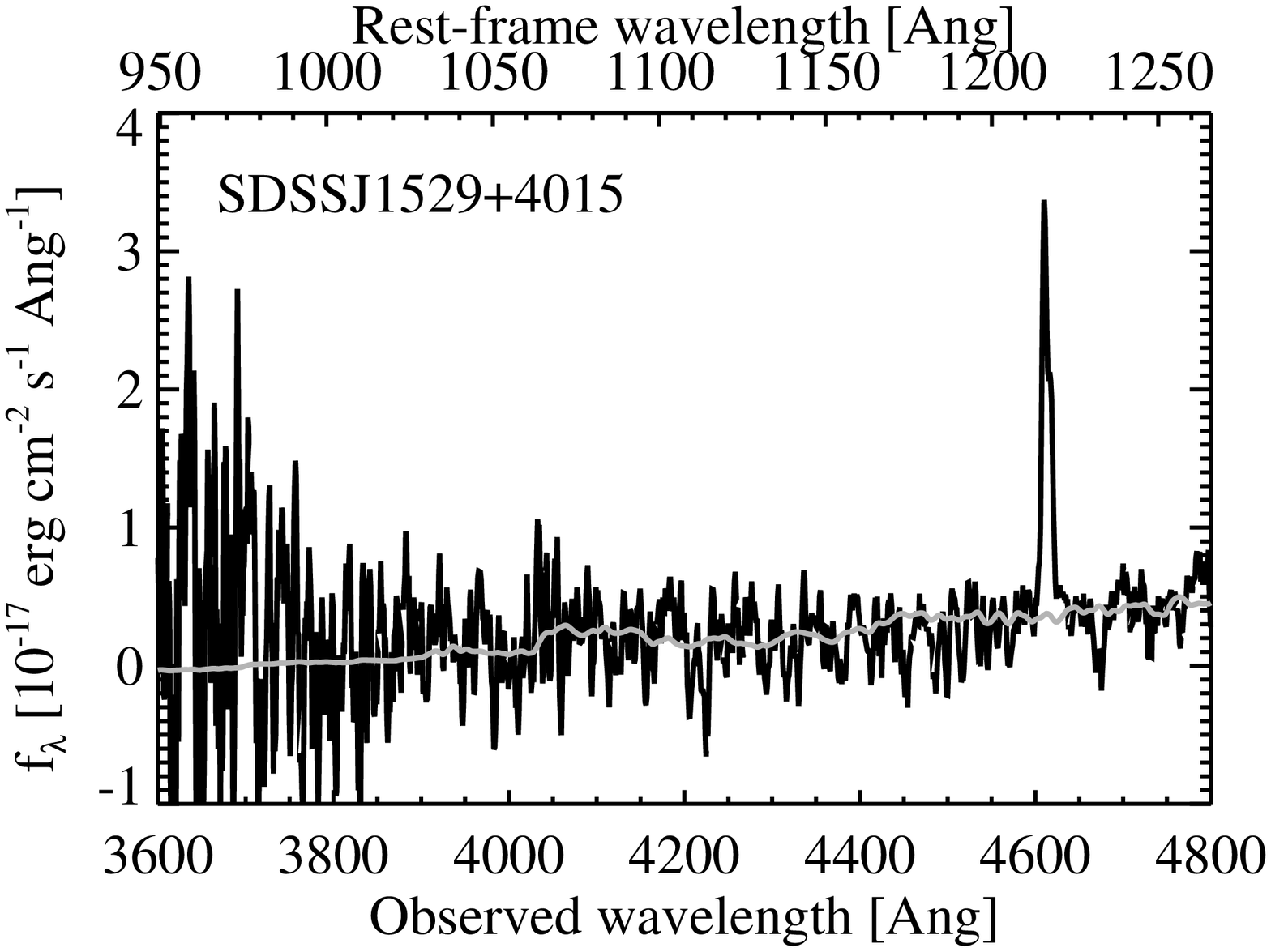}
\includegraphics[width=0.33\textwidth]{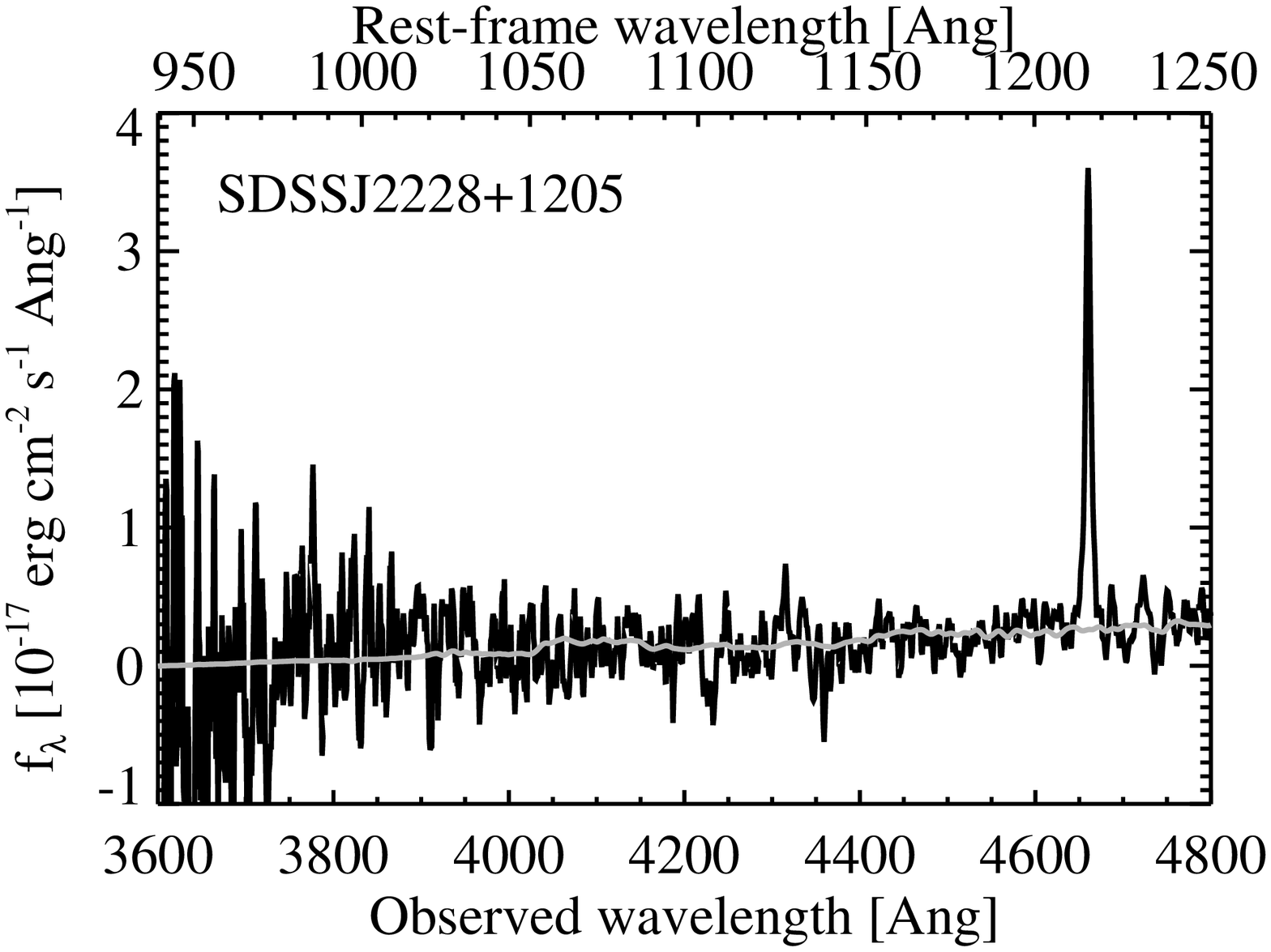}
\includegraphics[width=0.33\textwidth]{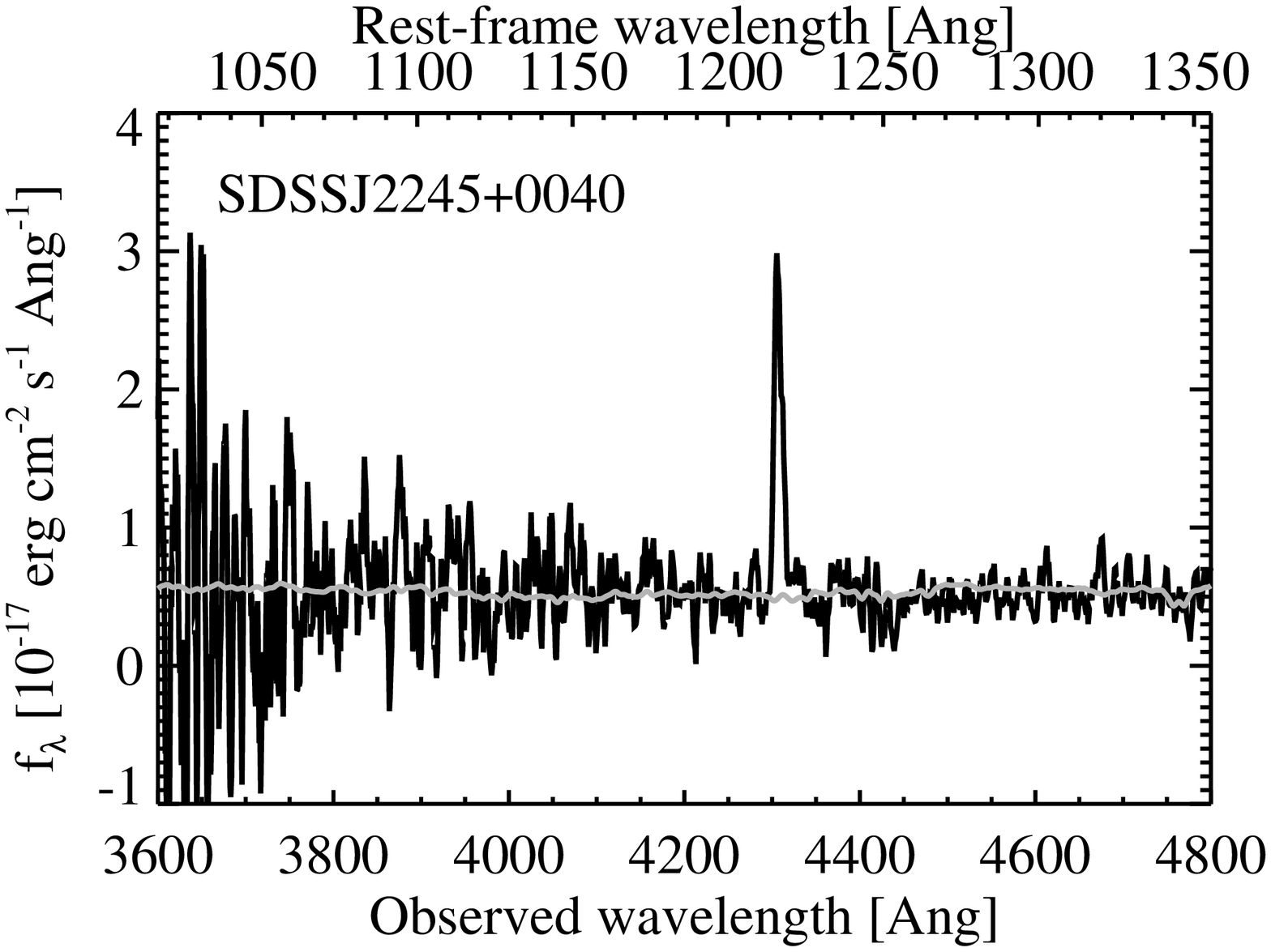}
\includegraphics[width=0.33\textwidth]{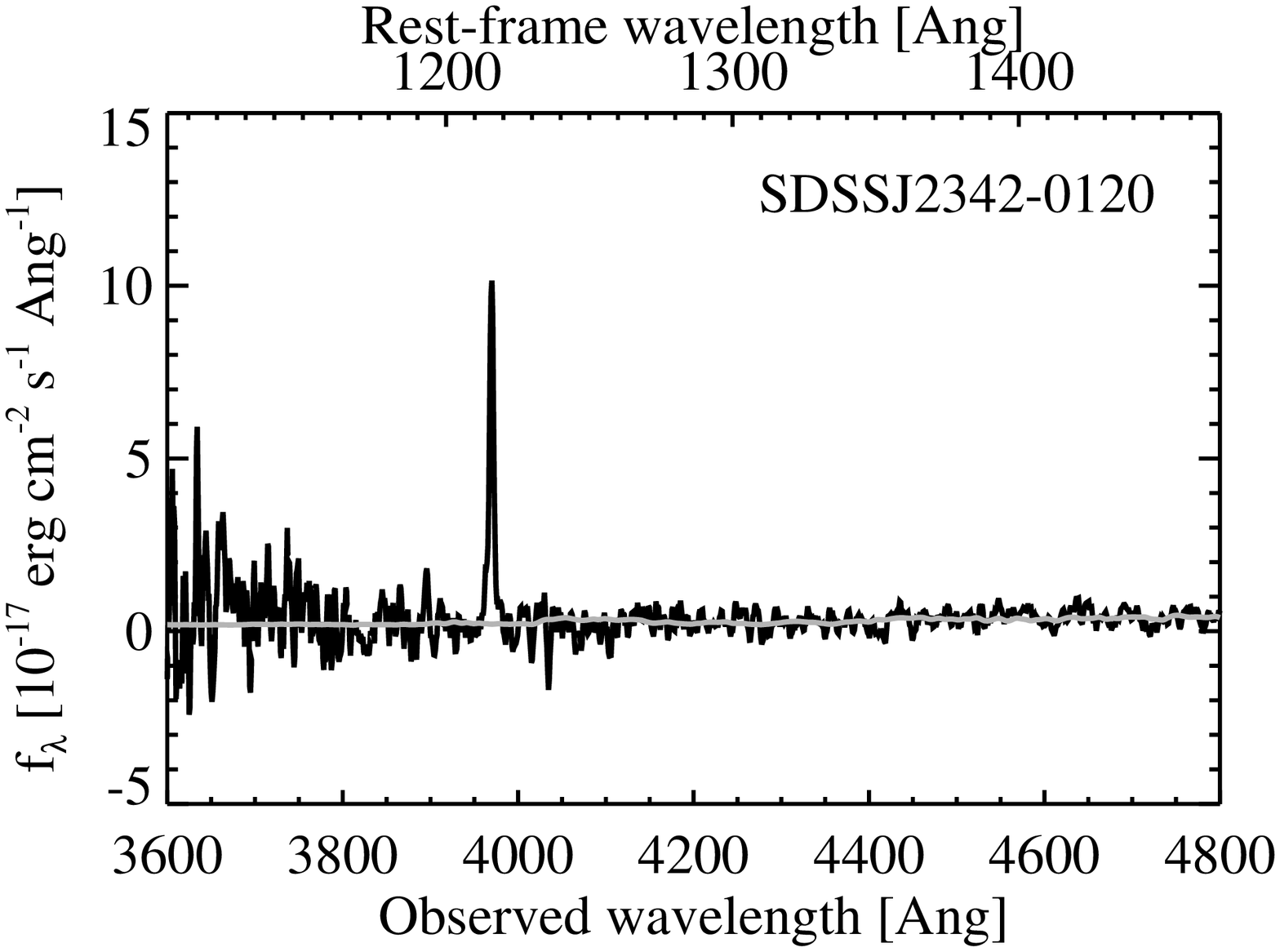}
\caption{{\label{fig:spectra2}} \textit{Continued}}
\end{figure*}
\begin{figure*}[htbp]
	\plotone{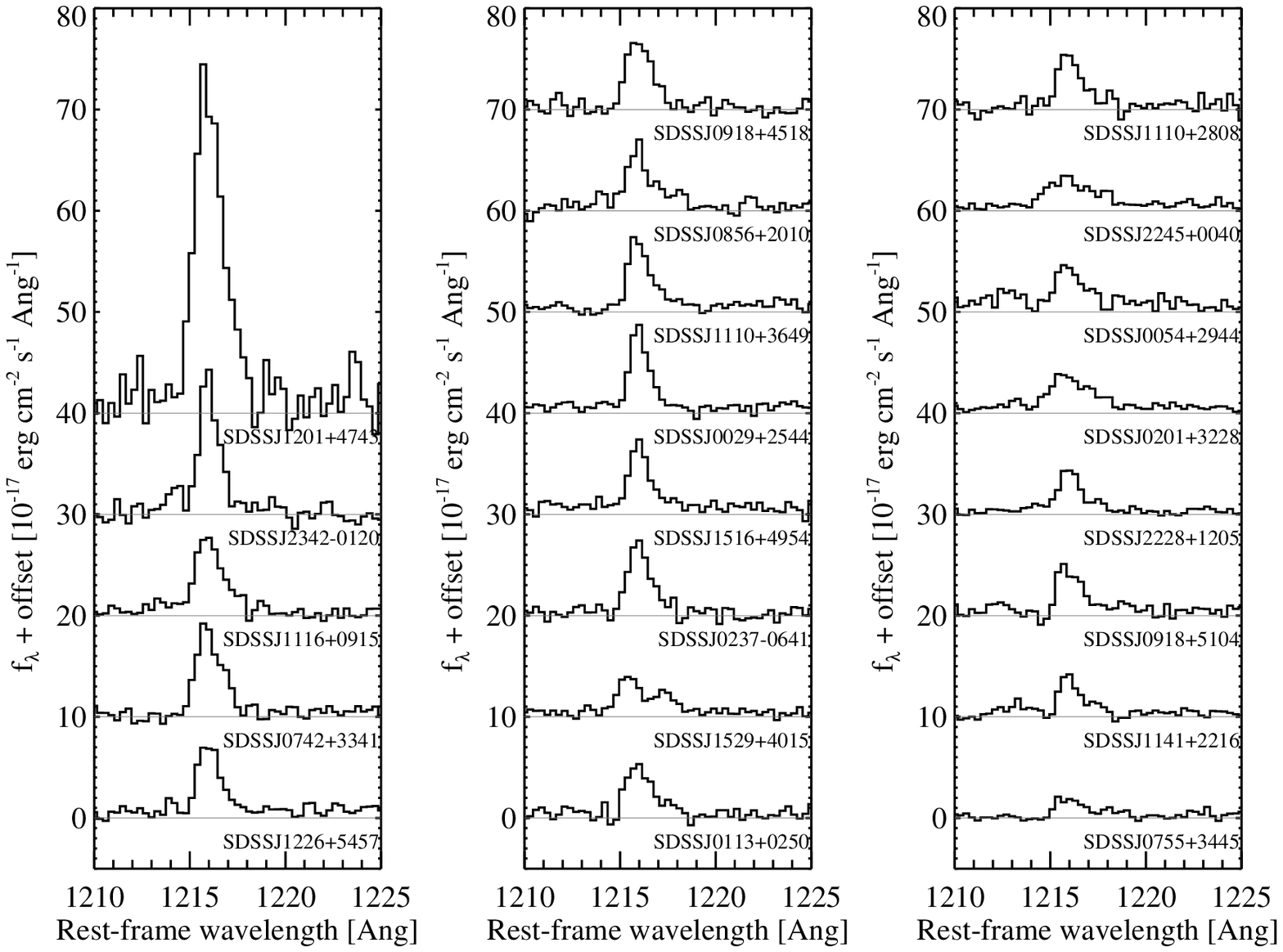}
\caption{\label{fig:lya}
Zoomed in views of the Ly$\alpha$ emission lines for the \Survey{} sample 
ordered by decreasing apparent Ly$\alpha$ flux, de-redshifted
to the rest frame of the LAEs. 
Gray lines are the best-fit continuum flux 
for the foreground galaxies. Note the ``blue edge, red tail'' 
line profiles characteristic of LAEs. }
\end{figure*}

\begin{figure*}[htbp]
	\plotone{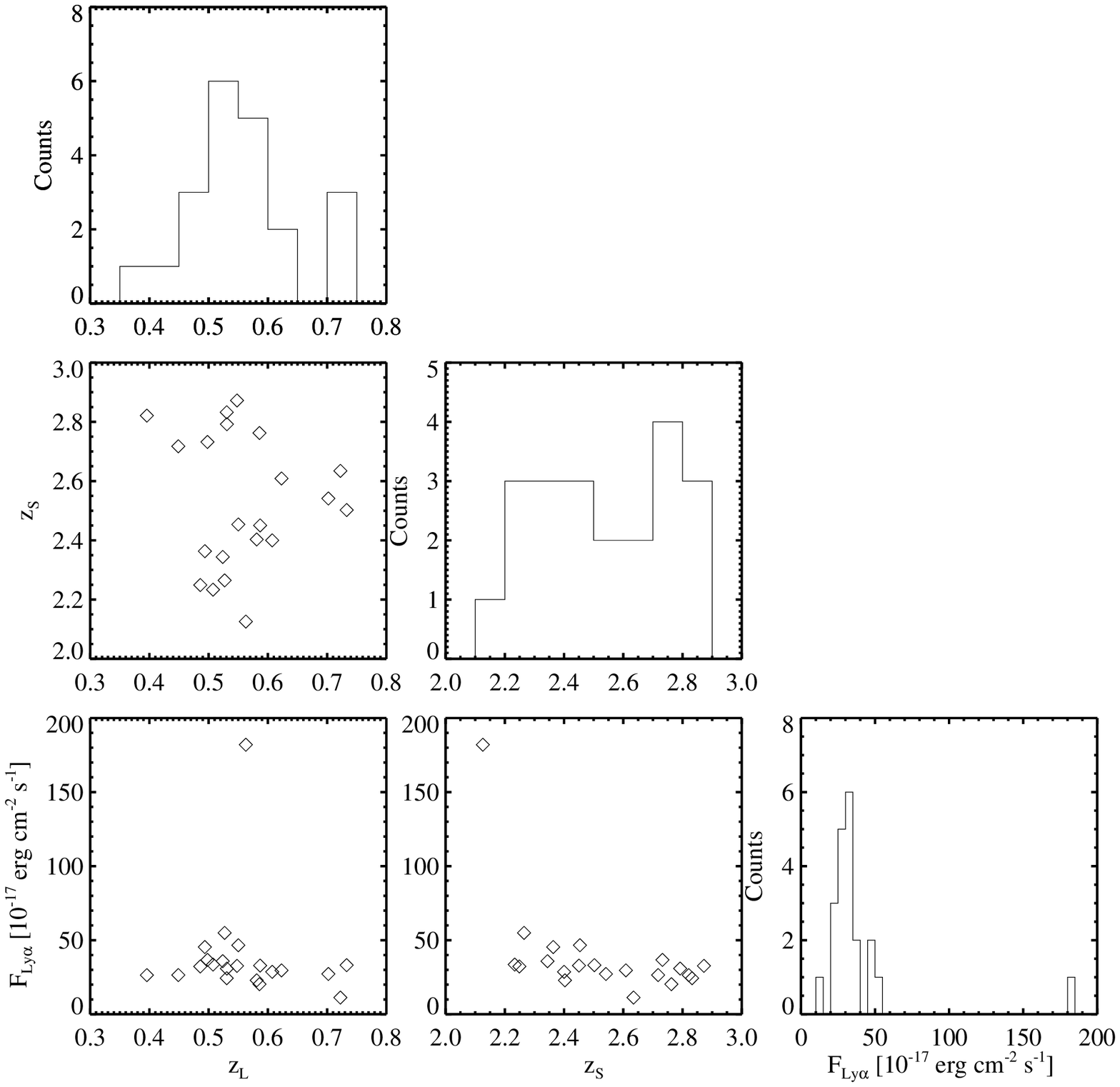}
\caption{\label{fig:diag}
Distributions of the \Survey{} sample in various parameter combinations. 
Histograms of the distributions are shown along the diagonal.}
\end{figure*}

\subsection{Spectroscopic and Photometric Properties}

All but two of our lens candidates are selected purely based on their spectroscopic data. 
Figure~\ref{fig:spectra1} displays the BOSS spectra (in black) 
observed within the 
$1{\arcsec}$-radius fibers for the 21 galaxy--LAE lens candidate 
systems, with a 5 pixel smoothing applied to suppress noise. 
We limit the spectra to the observed wavelength range from 3600\AA--4800\AA.  
Gray lines show the best-fit template for the continuum of 
the foreground BOSS galaxy as determined by the BOSS redshift and 
classification pipeline \citep{Bolton12b}. 
Since the foreground galaxies are early-type galaxies (ETGs) at redshift 
$z \sim 0.5$, there is little detectable foreground-galaxy flux within 
this wavelength range. 
All the continuum-subtracted residual spectra of these 21 systems 
exhibit strong emission features in the selected wavelength window.
We attribute the emission lines to the Ly$\alpha$ emission from high-redshift LAEs 
that happen to fall directly along the line of sight beyond the targeted BOSS 
galaxies since we have eliminated other possible line interlopers. 
Figure~\ref{fig:lya} presents a zoomed in view of the Ly$\alpha$ 
emission lines in the rest frame of the LAEs, ordered by 
decreasing apparent Ly$\alpha$ line flux. 
Note the``blue edge, red tail'' line profiles characteristic of LAEs. 

Table~\ref{tb:targets} reports basic spectroscopic and photometric 
properties of the 21 galaxy--LAE lens candidate systems as measured from SDSS/BOSS data.
The redshift of the foreground lens galaxy $z_L$ is measured spectroscopically 
as explained in \citet{Bolton12b}. 
The redshift of the lensed LAE $z_S$ can be inferred from the observed 
Ly$\alpha$ emission. The observed total flux of the 
Ly$\alpha$ emission $\rm F_{Ly \alpha}$ is estimated from the 
corresponding skew-normal fit and listed in units of 
$10^{-16}$\,erg\,cm$^{-2}$\,s$^{-1}$. As mentioned previously, two of the 21 
candidates (asterisked in the table) show definitive evidence for lensing features 
in their SDSS color images, while another two candidates show probable evidence for 
lensing features. 
More details about these four systems will be given in Section~\ref{sect:notes}. 
Table~\ref{tb:remaining_targets} in the Appendix summarizes the same 
spectroscopic and photometric properties for the remaining 166 galaxy--LAE 
lens candidate systems in the parent sample.

\begin{figure*}[htbp]
\centering
	\includegraphics[width=0.99\textwidth]{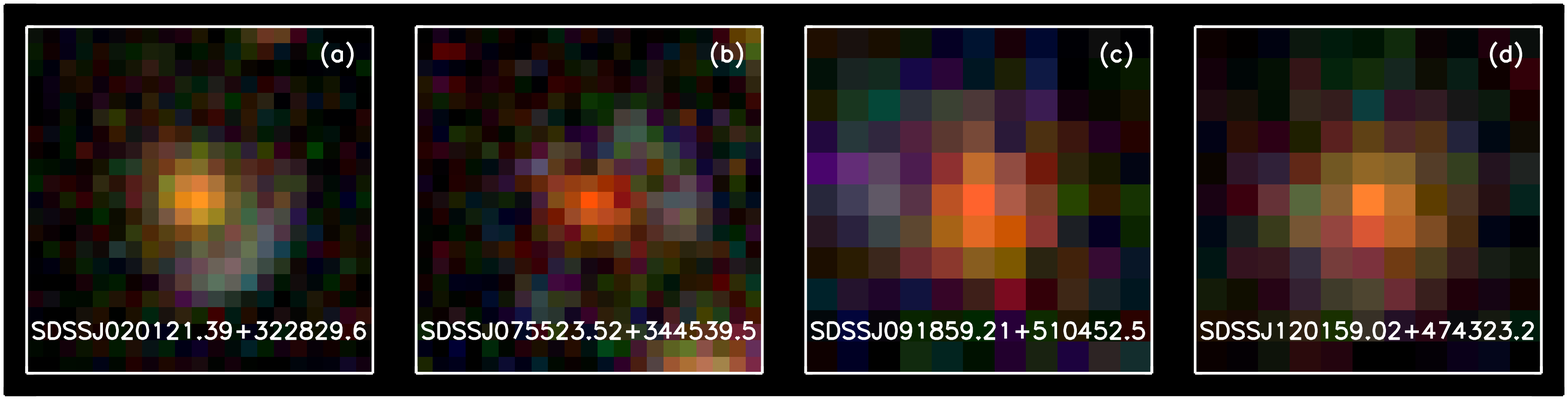}
\caption{\label{fig:mosaic}
Color-composite SDSS images of the four systems with strong (panels (a) and (b)) or 
probable (panels (c) and (d)) evidence for lensing signals. 
The first two panels are 4\arcsec$\times$4\arcsec, and the other two panels are 
2\arcsec$\times$2\arcsec. 
All images are orientated such that north is up and east is to the left. 
}
\end{figure*}

Figure~\ref{fig:diag} shows the distributions of the 21 galaxy--LAE 
lens candidate systems in the space of redshifts and Ly$\alpha$ flux. 
The redshift distribution of the foreground galaxies in our sample shows no 
significant deviation from that of the parent BOSS CMASS galaxy 
sample from which the candidates are selected \citep{Dawson13}. 
The source redshift distribution is relatively flat within $2.1 < z_S < 2.9$, 
and there is no correlation between $z_L$ and $z_S$. 
From the redshift distributions for both the foreground lens and background 
source, we do not see any significant systematic bias toward a particular 
redshift range. The observed total flux of the Ly$\alpha$ emission is generally 
between $2 \times 10^{-16}$\,erg\,cm$^{-2}$\,s$^{-1}$ and 
$5 \times 10^{-16}$\,erg\,cm$^{-2}$\,s$^{-1}$ except for one (SDSS\,J1201$+$4743) 
with an extremely high flux of $1.8 \times 10^{-15}$\,erg\,cm$^{-2}$\,s$^{-1}$. 
This highest-flux system also happens to have the closest source ($z_S = 2.1258$). 
As will be discussed in the following section, this system is one of the two
systems with probable strong-lensing features in their SDSS images. 
There is a mild correlation between the source redshift and the total apparent 
emission-line flux. Note that all flux values are reported here without any
correction for lensing magnification or fiber aperture losses.

\subsection{Notes on Individual Systems}
\label{sect:notes}

\emph {SDSS\,J020121.39$+$322829.6}---One of the two systems that exhibit definitive 
evidence for strong-lensing features in the corresponding color-composite 
SDSS images. This particular source shows a distinctive, extremely elongated arc with a 
relatively bluer color to the southwest of the foreground galaxy 
(panel (a) in Figure~\ref{fig:mosaic}). Although the 
foreground galaxy is a LOWZ instead of CMASS galaxy, the presence of the arc makes 
it an extremely promising candidate, and hence we include it in the \textsl{HST}
target list.

\emph{SDSS\,J075523.52$+$344539.5}---The other system that shows strong evidence 
for strong-lensing features in its color-composite SDSS image. 
Two distinctive bluish blobs connected by an arc are seen to the northwest 
and west of the foreground galaxy, with another faint blob to the southeast 
(panel (b) in Figure~\ref{fig:mosaic}). 
Although the detection significance 
and total flux of its emission line does not meet the selection thresholds, 
we include this system in the sample because of the suggestion of multiply lensed 
images.

\emph{SDSS\,J091859.21$+$510452.5}---A system that shows probable evidence 
for strong-lensing features in its color-composite SDSS image. A faint blue/greenish 
arc-like structure is seen to the northeast of the foreground galaxy
(panel (c) in Figure~\ref{fig:mosaic}).

\emph{SDSS\,J120159.02$+$474323.2}---The system with the highest apparent 
emission-line flux. Its color-composite SDSS image shows some greenish features 
surrounding the foreground galaxy as a possible indication of strong lensing 
(panel (d) in Figure~\ref{fig:mosaic}). 

\section{Strong-Lensing Probability}

For these systems to be powerful scientific tools, they must be bona fide 
strong lenses. Having applied several restrictive cuts on the 
significance, flux, and shape of the detected emission line, 
we are confident about the identification as high-redshift Ly$\alpha$.
In this section we estimate the probability that these LAEs are
strongly lensed.

To estimate the rate of strongly lensed LAE events within the sample, we simulate 
our spectroscopic selection procedures following the Monte Carlo approach of 
\citet{Arneson12}, with parameters tuned to the \Survey{} Survey. We model 
the LAE size distribution using exponential disks with a half-light radius 
drawn from an exponential distribution of $e$-folding scale 
$0\farcs05$ taken from \citet{Bond12}. 
We assume that the Ly$\alpha$ emission-line luminosity function has a simple 
power-law form as 
\begin{equation}
n(L, z) \propto L^{-\alpha}, 
\end{equation}
where $n(L, z)$ is the differential number density of LAEs at redshift $z$ 
with Ly$\alpha$ emission-line luminosity 
$L$ and $\alpha$ is the power-law index. 
We take $\alpha \simeq 1.5$ as an approximate average over 
the faint- and bright-end slopes in the redshift range 2.1--3.1
as studied by \citet{Ciardullo12}.
The simulation predicts that essentially all of our detected LAEs will be 
lensed into multiple images, and that 80\%, 50\%, and 20\% will have total 
magnifications exceeding 5, 10, and 30, respectively. 

As our candidates are selected largely based on the apparent emission-line flux, 
the magnification effect of gravitational lensing will lead to a so-called
magnification bias \citep[e.g.,][]{Gott74, Turner80, Vietri83}, 
in which the observed number 
counts of lensed sources in a flux-limited survey will be biased. For the 
assumed power-law luminosity function, the ratio of the source number counts 
between the lensed and unlensed scenarios can be estimated as
\begin{equation}
\frac{N(>L_0/\mu)}{N(>L_0)} = \mu^{\alpha-1}, 
\end{equation}
in which $N(>L_0)$ is the total number of LAEs with luminosity greater than 
$L_0$ and $\mu$ is the average magnification. Taking $\mu=10$ as a representative
value from our simulation, we would expect roughly a factor of $3.2$ boost in the 
number of lensed LAEs relative to the unlensed case. 

Given the typical velocity dispersion and redshift of a CMASS galaxy,
its multiple imaging cross section at the redshift of the LAE in our
sample will be comparable to or larger than the diameter of a
BOSS fiber. As a result, the multiply imaged source search region
associated with each CMASS galaxy is essentially the size of the
fiber. This is very different from general studies of lens statistics
\citep[e.g.,][]{Kochanek93a, Kochanek93b, Maoz93, Jaunsen95, Kochanek96, 
King96, Moller07} 
which must account for variations in the multiple imaging
cross section with lens redshift, lens mass, and source redshift. 

Given this fact, our estimate of the magnification bias can be checked 
observationally by comparing the incidence of LAE detections in BOSS galaxy 
spectroscopic fibers to their incidence within the blank-sky fibers of the survey. 
To explore this, we rerun the full candidate selection procedure on BOSS sky fibers 
and obtain 9 candidates with $> 8$-$\sigma$ detection significance. This number
can be compared to the 187 candidates found to this same significance
threshold in the galaxy-fiber sample. Since there are approximately
six times more galaxy fibers than sky fibers in the BOSS survey,
we find that our high-$z$ LAE detections occur with a frequency
$(187 / 9) / 6 \approx 3.5$ times higher in the galaxy fibers than
in the sky fibers. This is consistent with the boost factor expected based
on our simple magnification-bias calculation above: the higher relative
incidence of LAE detections in the galaxy fibers thus suggests that the
galaxy-fiber detections are highly magnified by strong lensing.
We stress, however, that the close quantitative agreement between the
predicted and observed detection-rate boost of galaxy fibers relative
to sky fibers must be regarded as coincidental considering the rough
approximations involved.

\section{Discussion}

We first emphasize that our spectroscopic selection technique is powerful and 
generic, and can be easily applied to a variety of scientific considerations with 
appropriate modifications. For instance, by changing the single-line detection 
strategy to detections for at least two of the lines 
[O\textsc{ii}]$\lambda \lambda$3727, [O\textsc{iii}]~4959,\,5007, 
H$\alpha$, and H$\beta$ at a common redshift, one can pick out lens candidates 
with star-forming galaxies as the background sources. That is exactly the approach 
adopted in the successful SLACS, BELLS, and S4TM surveys 
\citep{SLACSI, Brownstein2012, SLACSXII}. One can also specifically search for 
lensed quasars by tuning the selection criteria to the presence of more than one 
prominent quasar emission lines such as Ly$\alpha$, C \textsc{iv}, C \textsc{iii}], 
and Mg \textsc{ii}. Interested readers are encouraged to adopt the spectroscopic 
selection technique according to their own research interests.

Although the size of the sample of galaxy--LAE lenses from our initial 
\textsl{HST} program seems to be relatively modest 
when compared to other galaxy-scale strong-lens samples, 
there are another 166 galaxy--LAE lens candidates remaining in the parent sample. 
We applied very stringent requirements to pick the 21 highest-quality systems 
for this HST program, so many of the remaining 166 systems are also very good 
lens candidates. 
Considering the approximately 50\% success rates (the ratio of confirmed lenses 
to selected candidates) in previous spectroscopically selected lens samples 
in the SLACS, BELLS, and S4TM surveys 
\citep{SLACSV, Brownstein2012, SLACSXII}, we expect another $\sim 80$ 
galaxy-LAE lenses from further follow-ups of the \Survey{} parent sample. 
The scientific potential for such a large galaxy--LAE sample to constrain 
the nature and abundance of DM substructure is invaluable.

If the total mass--density profile of a lens galaxy and its
DM halo is modeled as a singular isothermal sphere, then the cut-off
mass above which substructure lensing effects are observable
scales as $m_c \propto \ell_s^{3/2}$, where $\ell_s$ is the characteristic
length scale of the lensed source \citep{Moustakas03}.
Previous works on LAEs between redshifts 2 and 3 have found a typical size of 
1--2 kpc in rest-frame UV light 
\citep[e.g.,][]{Bond10, Gronwall11, McLinden11, Law12, Malhotra12, Ryan12, Bond12}. 
More recently, analysis by \citet{Shu16} of high-resolution
\textsl{HST} imaging of the lensed $z = 2.7$
LAE system originally described by \citet{Bolton06b}
suggests that the background LAE can even be resolved into three distinct 
components with characteristic sizes from 116--438 pc. 
By comparison, the typical physical size of the sources in
previously assembled samples of strong galaxy--galaxy lenses
(e.g., SLACS, BELLS, and S4TM) is 1-2 kpc \citep[e.g.,][]{SLACSXI}.
Thus, these new galaxy--LAE lens systems have the potential to 
push the lensing detection limit of DM substructure
lower by an order of magnitude. 

Certainly, the actual DM substructure detectability depends 
strongly on the instrumental resolution, sensitivity, and the S/N. 
Observations in radio/near UV or with the aid of adaptive optics, which in general 
offer higher resolutions and/or S/Ns relative to the HST observations of 
the \Survey{} sample, are also important for DM substructure detection 
\citep[e.g.,][]{Bradavc02, Dalal02, Vegetti12, MacLeod13, 
Nierenberg14, Hezaveh16, Inoue16b}. 
Nevertheless, the \Survey{} lens sample itself is invaluable because of 
its sensitivity to lower-mass substructures than the current 
strong galaxy--galaxy lenses. 

The benefit from this factor of $\sim 10$ or more boost in the substructure-mass 
detection limit is threefold. First, lower-mass substructures are much more abundant 
($\sim \times 100$) according to the predicted CDM subhalo mass function, which is found to 
approximately follow a fast-declining power-law profile in semi-analytical models 
and high-resolution cosmological simulations 
\citep[e.g.,][]{Bullock01, DeLucia04, Gao04, vandenBosch05, Zentner05, Giocoli08}. 
Hence the rate of substructure detection should be higher in this
sample than in previous strong-lens samples.
Second, sensitivity over a wider range of substructure-masses can lead to
constraints on the slope of the substructure mass function
in addition to the overall abundance at a given mass.
Third, this new sample can probe for the first time the regime in which 
the predictions of competing DM models differ significantly 
\citep[e.g.,][]{Colin00, Spergel00, AvilaReese01, Springel08, 
Lovell12, Lovell14, Li15, Bose16}.

Considering the positive detection of mass substructures in 1/11 galaxy--galaxy 
lens systems in the \citet{Vegetti14} analysis, 
our proposed sample of 21 galaxy--LAE lens systems can be expected to show  
strong evidence for dark substructure even allowing for the uncertainties 
in forecasting. 
We therefore expect this \Survey{} program with 21 galaxy--LAE lens systems 
to provide the most precise measurement to date of the substructure-mass 
function in galaxy-scale halos. Alternatively, 
the non-detection of significant substructure within this new lensed LAE 
sample would provide surprising evidence for the suppression of 
low-mass substructure below an unexpectedly high cut-off mass.

A complementary study enabled by this unique sample of galaxy--LAE lens systems 
is of the mass--density profile of massive ETGs. The foreground lens population 
in our sample is comparable in all ways to the earlier sample of BELLS lenses, 
but the higher source redshifts of the LAE lens sample will lead to the formation
of lensed images with larger angular Einstein radii. 
The typical source redshift of BELLS galaxy--galaxy lenses is $\sim 1$. 
Considering a typical BOSS lens galaxy at $z=0.52$, a $z \sim 3$ LAE will 
hence correspond to an Einstein radius that is larger by a factor of 1.7. Since 
gravitational lensing provides a precise mass estimation within the Einstein radius, 
a combined ensemble analysis of both samples with a range of Einstein radii 
can provide more precise constraints on the radial mass--density profile in massive 
ETGs and its decomposition into stars and DM. 
Other strong-lens programs such as the SDSS Quasar Lens 
Search \citep{Inada12} and the Strong Lensing Legacy Survey 
\citep{Sonnenfeld13a} have also found lenses with similar lens and source 
redshifts. However, their lens selection functions and hence lens populations are 
different from the BELLS survey, and thereby cannot be directly combined with 
BELLS lenses as the \Survey{} lenses can. 

In addition to the explorations of the foreground lens galaxies, this new sample 
will enable other studies of high-redshift LAEs. Direct observations of 
high-redshift LAEs are severely limited by S/N and resolution and 
therefore strongly biased toward the luminous end. 
In many cases a stacking technique 
must be adopted to compensate for low photon counts, and can only provide a 
rough picture of the entire LAE sample \citep[e.g.,][]{Steidel11, Matsuda12, 
Feldmeier13}. Exploration of detailed structures of individual LAEs is challenging 
and sometimes impossible. Strong gravitational lensing aids in the study of 
high-redshift LAEs through its magnification effect, in which the apparent size of 
the lensed source is magnified up to factors of tens 
\citep[e.g.,][]{Bolton06b, Bayliss10, Christensen12, Jones13}. 
Therefore, in combination with the unmatched spatial resolution of
\textsl{HST}, lensing will allow us to explore the details of LAEs 
below 100 pc scales and to see a clearer picture of the sites of the most 
concentrated unobscured star formation in the universe. 
Additionally, as gravitational lensing preserves source surface brightness, 
the magnification in size leads to an equal boost in the photon counts, 
which significantly improves the detectability and precision in flux measurement. 
This boost also permits the exploration of intrinsically faint 
LAEs which would otherwise require large investments of time in blank-field searches.
Once the detailed selection function of our lensed LAE survey is understood through 
\textsl{HST} imaging and spatially resolved ground-based spectroscopy, 
the full sample of LAEs detected in BOSS spectra can be used for a measurement of 
the joint size-luminosity function of LAEs and its evolution from $z = 3$ to $2$. 

Our program will have important implications as a pathfinder for 
more ambitious lensed LAE surveys within BOSS, and for future spectroscopic 
projects such as DESI \citep{Flaugher14} and Subaru-PFS \citep{Sugai12}. 
Finally, based on the example of SLACS and other samples of significantly 
new astronomical systems, numerous unanticipated scientific applications 
will no doubt arise on the basis of this new lens sample.

\section{Conclusion}

We have presented a catalog of 21 galaxy--LAE lens candidate systems as 
a part of the \Survey{} Survey, the goal of which is to ``illuminate'' dark 
substructures in galaxy-scale systems and deliver new constraints on the nature 
of DM and its dynamics. The selection strategy is analogous to the technique adopted 
in the successful SLACS, BELLS, and S4TM surveys,
modified to detect lensed high-redshift LAEs as the background sources. 
Benefiting from the compact nature of these LAEs, this new sample
has the potential to push the lensing detection limit 
of DM substructure down by an order of magnitude in terms of its mass 
relative to currently existing galaxy--galaxy lens systems. 

The parent sample of 187 galaxy--LAE candidates is selected spectroscopically 
from the BOSS final data product (DR12) with $\approx 1.4$ million galaxy optical spectra 
by searching for the existence of high-significance Ly$\alpha$ emission 
in the observed wavelength range of 3600\AA~$< \lambda <$4800\,\AA. 
We also find nine emission-line detections in the BOSS blank sky fibers with the 
same selection procedures. Such detection enhancement within galaxy fibers is 
consistent with the lensing magnification-bias effect and the prediction from Monte 
Carlo simulations that suggest a very high strong-lensing rate in the candidate 
sample.

By applying multiple quality cuts, we build a final sample of 21 
highest-quality galaxy--LAE lens candidate systems for \textsl{HST} 
high-resolution follow-up imaging, which is being carried out under
\textsl{HST} Cycle 23 GO Program \# 14189 (PI: A. S. Bolton). Four of the 
21 candidates already exhibit either definitive or probable evidence for 
lensing features (elongated arc and multiple images) in their SDSS images. 
The redshift distribution of the lens galaxies is roughly Gaussian, centered 
at $z \sim 0.5$, with no significant difference from the distribution
of redshifts in the parent BOSS CMASS galaxy sample
(one candidate is actually a LOWZ galaxy). 
The background LAEs have a relatively flat redshift distribution
from $z=2.1$ to 2.9, suggesting little selection bias toward any redshift range. 

In consideration of a previous detection of dark substructure
in the S/N-selected SLACS galaxy--galaxy lens sample \citep{Vegetti10}, 
we expect the \Survey{} program and similar programs in the future
to illuminate dark substructure
within galaxy-scale halos, to provide a precise observational measurement
of the substructure-mass function observationally, and to combine with
theoretical and numerical predictions to lead to a better understanding
of the nature of DM.

Finally, for interested readers, the source code (in Interactive Data Language) 
implementing the described selection 
technique can be downloaded from the Bitbucket repository at 
\url {https://bitbucket.org/abolton/lae_lens}. Installation of \texttt{IDLUTLIS}, 
a collection of IDL utilities, is also required and can be done following 
\url {http://www.sdss.org/dr12/software/idlutils/}. 
Our code works specifically with BOSS spectroscopic data which are organized in 
a particular plate--MJD-fiber structure. Nevertheless, the kernel of the code, 
the spectroscopic selection technique, is generic and can be easily adopted to 
data sets in different structures with moderate modifications.

\acknowledgments

This work has been partially supported by the Strategic Priority Research Program ``The Emergence of Cosmological Structures'' of the Chinese Academy of Sciences Grant No. XDB09000000 and by the National Natural Science Foundation of China (NSFC) under grant numbers 11333003 and 11390372 (Y.S. and S.M.). The support and resources from the Center for High Performance Computing at the University of Utah are gratefully acknowledged. The work of M.O. was supported in part by World Premier International Research Center Initiative (WPI Initiative), MEXT, Japan, and JSPS KAKENHI Grant Numbers 26800093 and 15H05892. Z.Z. is partially supported by NASA grant NNX14AC89G and NSF grant AST-1208891. 

Funding for SDSS-III was provided by the Alfred P. Sloan Foundation, the Participating Institutions, the National Science Foundation, and the U.S. Department of Energy Office of Science. The SDSS-III web site is http://www.sdss3.org/.

SDSS-III was managed by the Astrophysical Research Consortium for the Participating Institutions of the SDSS-III Collaboration including the University of Arizona, the Brazilian Participation Group, Brookhaven National Laboratory, Carnegie Mellon University, University of Florida, the French Participation Group, the German Participation Group, Harvard University, the Instituto de Astrofisica de Canarias, the Michigan State/Notre Dame/JINA Participation Group, Johns Hopkins University, Lawrence Berkeley National Laboratory, Max Planck Institute for Astrophysics, Max Planck Institute for Extraterrestrial Physics, New Mexico State University, New York University, Ohio State University, Pennsylvania State University, University of Portsmouth, Princeton University, the Spanish Participation Group, University of Tokyo, University of Utah, Vanderbilt University, University of Virginia, University of Washington, and Yale University.

\bibliographystyle{apj}

\appendix
\begin{table*}[htbp]
\begin{center}
\caption{\label{tb:remaining_targets} Selected properties of the remaining 166 galaxy--LAE lens candidate system in the \Survey{} parent sample.}
\begin{tabular}{l c c c c c c c}
\hline \hline
Target & Plate--MJD--Fiber & $z_{L}$ & $z_{s}$ & R.A. & Decl. & $m_i$ & Ly$\alpha$ Flux \\
\hline
SDSSJ001213.49$+$173537.5 & 6111-56270-279 & 0.4911 & 1.9886 & 00 12 13.4911 & $+$17 35 37.5687 & 19.06 & 53.86 \\
\hline
SDSSJ001620.58$-$061115.5 & 7150-56597-586 & 0.3015 & 2.3111 & 00 16 20.5816 & $-$06 11 15.5836 & 17.74 & 22.80 \\
\hline
SDSSJ002123.23$-$042926.1 & 7037-56570-769 & 0.5583 & 2.1453 & 00 21 23.2327 & $-$04 29 26.1655 & 19.77 & 24.14 \\
\hline
SDSSJ003338.97$+$042734.9 & 4418-55862-165 & 0.5264 & 2.1854 & 00 33 38.9738 & $+$04 27 34.9358 & 19.66 & 25.30 \\
\hline
SDSSJ005753.31$+$001600.7 & 4224-55481-929 & 0.4937 & 2.6005 & 00 57 53.3192 & $+$00 16 00.6960 & 19.19 & 23.92 \\
\hline
SDSSJ005948.32$+$160629.4 & 5705-56194-259 & 0.5946 & 2.8324 & 00 59 48.3202 & $+$16 06 29.4928 & 20.01 & 10.43 \\
\hline
SDSSJ010701.12$-$030830.6 & 4373-55811-336 & 0.2452 & 2.7271 & 01 07 01.1261 & $-$03 08 30.6677 & 17.61 & 19.60 \\
\hline
SDSSJ011120.66$-$031559.4 & 4373-55811- 88 & 0.4480 & 2.9289 & 01 11 20.6648 & $-$03 15 59.4154 & 18.74 & 13.53 \\
\hline
SDSSJ012233.11$-$052239.9 & 7046-56568-101 & 0.6539 & 2.2641 & 01 22 33.1169 & $-$05 22 39.9855 & 19.80 & 13.12 \\
\hline
SDSSJ013148.59$-$053037.6 & 7048-56575-335 & 0.4873 & 2.7885 & 01 31 48.5934 & $-$05 30 37.6247 & 19.86 & 15.61 \\
\hline
SDSSJ013644.87$-$064656.6 & 7162-56605-731 & 0.6686 & 2.9470 & 01 36 44.8787 & $-$06 46 56.6904 & 19.90 & 11.74 \\
\hline
SDSSJ013914.73$+$044334.6 & 4274-55508-850 & 0.2437 & 2.3991 & 01 39 14.7350 & $+$04 43 34.6793 & 17.30 & 31.43 \\
\hline
SDSSJ014526.41$-$050035.5 & 7050-56573-273 & 0.5696 & 2.7219 & 01 45 26.4148 & $-$05 00 35.5906 & 19.95 & 15.16 \\
\hline
SDSSJ014631.97$-$012514.6 & 4350-55556-945 & 0.5213 & 2.6038 & 01 46 31.9762 & $-$01 25 14.6048 & 19.70 & 12.88 \\
\hline
SDSSJ015138.11$+$005559.1 & 4233-55449-587 & 0.4857 & 2.8007 & 01 51 38.1175 & $+$00 55 59.1941 & 18.87 & 9.38 \\
\hline
SDSSJ015145.35$-$0000-1.3 & 3606-55182-695 & 0.4901 & 2.3563 & 01 51 45.3552 & $-$00 00 -1.3063 & 19.00 & 11.47 \\
\hline
SDSSJ015150.88$+$143037.3 & 4657-55591-602 & 0.5623 & 2.2461 & 01 51 50.8855 & $+$14 30 37.3569 & 19.18 & 26.54 \\
\hline
SDSSJ020103.73$+$161130.8 & 5119-55836-846 & 0.1110 & 2.9099 & 02 01 03.7308 & $+$16 11 30.8203 & 16.13 & 16.03 \\
\hline
SDSSJ020241.40$-$064611.3 & 4398-55946-379 & 0.5020 & 2.7477 & 02 02 41.4084 & $-$06 46 11.3322 & 19.40 & 12.55 \\
\hline
SDSSJ021742.02$-$002206.4 & 4236-55479-117 & 0.6069 & 2.4791 & 02 17 42.0209 & $-$00 22 06.4687 & 19.46 & 13.00 \\
\hline
SDSSJ074502.39$+$181601.9 & 4488-55571- 49 & 0.5397 & 2.8723 & 07 45 02.3969 & $+$18 16 01.9423 & 19.07 & 13.30 \\
\hline
SDSSJ074720.77$+$272914.9 & 4452-55536-205 & 0.2713 & 2.1796 & 07 47 20.7788 & $+$27 29 14.9698 & 17.43 & 39.15 \\
\hline
SDSSJ074816.73$+$445602.8 & 3676-55186-623 & 0.5394 & 2.1854 & 07 48 16.7377 & $+$44 56 02.8326 & 19.06 & 30.10 \\
\hline
SDSSJ074831.92$+$471014.4 & 3673-55178-795 & 0.4387 & 2.5056 & 07 48 31.9208 & $+$47 10 14.4104 & 19.64 & 16.30 \\
\hline
SDSSJ075057.45$+$221035.8 & 4473-55589-231 & 0.4516 & 2.7564 & 07 50 57.4512 & $+$22 10 35.8337 & 19.33 & 13.30 \\
\hline
SDSSJ075902.52$+$441705.0 & 3683-55178-662 & 0.4670 & 2.8857 & 07 59 02.5287 & $+$44 17 05.0656 & 19.76 & 11.09 \\
\hline
SDSSJ075930.24$+$441245.5 & 3683-55245-671 & 0.4668 & 2.8857 & 07 59 30.2490 & $+$44 12 45.5548 & 20.04 & 9.00 \\
\hline
SDSSJ081535.86$+$452133.5 & 3691-55274-138 & 0.5642 & 2.3773 & 08 15 35.8612 & $+$45 21 33.5715 & 20.12 & 22.74 \\
\hline
SDSSJ081627.70$+$530140.5 & 3690-55182-643 & 0.5333 & 2.4243 & 08 16 27.7057 & $+$53 01 40.5112 & 19.47 & 17.00 \\
\hline
SDSSJ082350.26$+$425114.1 & 3807-55511-153 & 0.6034 & 2.2890 & 08 23 50.2606 & $+$42 51 14.1669 & 19.82 & 14.05 \\
\hline
SDSSJ082433.59$+$190704.5 & 4491-55570-827 & 0.5213 & 2.8821 & 08 24 33.5980 & $+$19 07 04.5529 & 19.74 & 12.31 \\
\hline
SDSSJ083023.30$+$270017.9 & 4460-55533-583 & 0.5201 & 2.4235 & 08 30 23.3057 & $+$27 00 17.9008 & 19.46 & 21.27 \\
\hline
SDSSJ083052.16$+$212858.1 & 4478-55600-135 & 0.6195 & 2.2626 & 08 30 52.1667 & $+$21 28 58.1332 & 19.91 & 50.65 \\
\hline
SDSSJ083324.57$+$151215.4 & 4496-55544- 93 & 0.4501 & 1.9907 & 08 33 24.5728 & $+$15 12 15.4488 & 19.57 & 54.92 \\
\hline
SDSSJ084211.29$+$383125.6 & 3765-55508-727 & 0.5816 & 2.8492 & 08 42 11.2903 & $+$38 31 25.6934 & 19.57 & 61.33 \\
\hline
SDSSJ084246.17$+$362456.1 & 4609-56251-483 & 0.5169 & 2.8191 & 08 42 46.1719 & $+$36 24 56.1044 & 19.57 & 16.90 \\
\hline
SDSSJ084300.13$+$194742.6 & 5176-56221-583 & 0.6985 & 2.5462 & 08 43 00.1392 & $+$19 47 42.6251 & 19.81 & 16.72 \\
\hline
SDSSJ084443.21$+$012504.3 & 3810-55531-231 & 0.5911 & 2.7391 & 08 44 43.2129 & $+$01 25 04.3223 & 20.00 & 8.58 \\
\hline
SDSSJ090254.78$-$005745.8 & 3818-55532-409 & 0.5786 & 2.8359 & 09 02 54.7888 & $-$00 57 45.8659 & 19.43 & 9.08 \\
\hline
SDSSJ091526.15$+$585155.3 & 5712-56602-365 & 0.7812 & 2.8244 & 09 15 26.1548 & $+$58 51 55.3244 & 19.83 & 13.72 \\
\hline
SDSSJ091727.90$+$250027.4 & 5777-56280-230 & 0.3261 & 2.4401 & 09 17 27.9089 & $+$25 00 27.4658 & 18.19 & 13.36 \\
\hline
SDSSJ092133.07$+$363717.4 & 4644-55922-885 & 0.5558 & 2.6819 & 09 21 33.0762 & $+$36 37 17.4069 & 19.91 & 15.66 \\
\hline
SDSSJ092159.66$+$070014.5 & 4870-55923-799 & 0.5903 & 2.2349 & 09 21 59.6631 & $+$07 00 14.5483 & 19.62 & 17.31 \\
\hline
SDSSJ092335.14$+$190554.7 & 5767-56245-437 & 0.4387 & 2.5113 & 09 23 35.1416 & $+$19 05 54.7485 & 18.45 & 14.97 \\
\hline
SDSSJ093551.05$+$085130.3 & 5314-55952-733 & 0.4569 & 2.5601 & 09 35 51.0498 & $+$08 51 30.3065 & 19.55 & 18.46 \\
\hline
SDSSJ093741.40$-$013203.8 & 3767-55214-925 & 0.5094 & 2.9334 & 09 37 41.4038 & $-$01 32 03.8081 & 19.70 & 16.27 \\
\hline
SDSSJ100149.69$+$273647.1 & 6471-56309-355 & 0.6218 & 2.1957 & 10 01 49.6985 & $+$27 36 47.1602 & 19.61 & 14.55 \\
\hline
SDSSJ101908.39$+$250436.5 & 6465-56279-275 & 0.6473 & 2.4655 & 10 19 08.3936 & $+$25 04 36.5190 & 19.09 & 13.18 \\
\hline
SDSSJ102529.98$+$424926.0 & 4557-55588-812 & 0.3413 & 2.6196 & 10 25 29.9890 & $+$42 49 26.0339 & 17.75 & 11.08 \\
\hline
SDSSJ102901.22$+$472822.6 & 6659-56607-751 & 0.5696 & 2.8288 & 10 29 01.2268 & $+$47 28 22.6337 & 19.11 & 19.70 \\
\hline
SDSSJ102938.39$+$154743.8 & 5340-56011-901 & 0.5607 & 2.1584 & 10 29 38.3899 & $+$15 47 43.8268 & 19.61 & 34.94 \\
\hline
SDSSJ103117.27$+$262354.9 & 6457-56330-371 & 0.5478 & 2.7859 & 10 31 17.2778 & $+$26 23 54.9586 & 19.68 & 11.49 \\
\hline
SDSSJ105018.46$+$090256.4 & 5354-55927-721 & 0.6180 & 2.4839 & 10 50 18.4680 & $+$09 02 56.4404 & 19.56 & 12.61 \\
\hline
SDSSJ105222.56$+$472252.7 & 6664-56383-669 & 0.4747 & 2.2038 & 10 52 22.5659 & $+$47 22 52.7417 & 18.96 & 11.57 \\
\hline
SDSSJ105507.57$+$314126.4 & 6445-56366-875 & 0.5189 & 2.4195 & 10 55 07.5769 & $+$31 41 26.4670 & 19.93 & 18.60 \\
\hline
SDSSJ105620.82$+$114637.0 & 5356-55979-743 & 0.4698 & 2.2951 & 10 56 20.8228 & $+$11 46 37.0264 & 19.90 & 15.66 \\
\hline
SDSSJ105752.35$+$064431.2 & 4854-55685-843 & 0.5961 & 2.9108 & 10 57 52.3572 & $+$06 44 31.2948 & 19.44 & 10.59 \\
\hline
SDSSJ105909.96$+$110215.1 & 5356-55979-885 & 0.5459 & 2.6221 & 10 59 09.9609 & $+$11 02 15.1524 & 18.89 & 18.03 \\
\hline
SDSSJ110148.19$+$523918.9 & 6706-56385-855 & 0.5821 & 2.8042 & 11 01 48.1970 & $+$52 39 18.9294 & 19.87 & 8.05 \\
\hline
SDSSJ110502.65$+$035223.8 & 4771-55925-113 & 0.4929 & 2.7867 & 11 05 02.6514 & $+$03 52 23.8632 & 19.54 & 7.40 \\
\hline \hline
\end{tabular}
\end{center}
\end{table*}
\addtocounter{table}{-1}
\begin{table*}[htbp]
\begin{center}
\caption{\textit{Continued}}
\begin{tabular}{l c c c c c c c}
\hline \hline
Target & Plate--MJD--Fiber & $z_{L}$ & $z_{s}$ & R.A. & Decl. & $m_i$ & Ly$\alpha$ Flux \\
\hline
SDSSJ111152.93$+$381222.2 & 4621-55649-257 & 0.5274 & 2.7503 & 11 11 52.9395 & $+$38 12 22.2775 & 19.53 & 9.54 \\
\hline
SDSSJ111544.57$+$084717.1 & 5366-55958-259 & 0.5523 & 2.8902 & 11 15 44.5715 & $+$08 47 17.1574 & 19.54 & 15.83 \\
\hline
SDSSJ111945.18$+$441146.0 & 6648-56383-253 & 0.6979 & 2.3540 & 11 19 45.1831 & $+$44 11 46.0776 & 19.88 & 14.96 \\
\hline
SDSSJ112159.89$+$525327.3 & 6698-56637-387 & 0.4655 & 2.1671 & 11 21 59.8938 & $+$52 53 27.3486 & 19.87 & 35.11 \\
\hline
SDSSJ112327.36$+$305352.6 & 6434-56362-705 & 0.4306 & 2.5137 & 11 23 27.3669 & $+$30 53 52.6515 & 18.59 & 15.75 \\
\hline
SDSSJ112429.08$+$583124.9 & 7100-56636-812 & 0.1960 & 2.3874 & 11 24 29.0881 & $+$58 31 24.9518 & 17.60 & 12.78 \\
\hline
SDSSJ112558.59$+$335647.1 & 4619-55599-680 & 0.4824 & 2.9216 & 11 25 58.5901 & $+$33 56 47.1075 & 19.46 & 13.59 \\
\hline
SDSSJ112708.12$+$652920.2 & 7110-56746-947 & 0.5358 & 2.6473 & 11 27 08.1226 & $+$65 29 20.2020 & 19.48 & 14.23 \\
\hline
SDSSJ113110.97$+$355020.7 & 4615-55618-399 & 0.4685 & 2.2974 & 11 31 10.9790 & $+$35 50 20.7184 & 19.77 & 34.76 \\
\hline
SDSSJ113938.99$+$165158.5 & 5891-56034-104 & 0.2791 & 2.4751 & 11 39 38.9978 & $+$16 51 58.5242 & 17.89 & 46.45 \\
\hline
SDSSJ114358.81$+$505552.1 & 6684-56412-835 & 0.5236 & 2.1737 & 11 43 58.8135 & $+$50 55 52.1622 & 19.24 & 40.12 \\
\hline
SDSSJ115136.39$+$233812.8 & 6423-56313-755 & 0.6404 & 2.9425 & 11 51 36.3977 & $+$23 38 12.8398 & 19.94 & 9.50 \\
\hline
SDSSJ115422.81$-$025127.8 & 3776-55209-451 & 0.5628 & 2.9018 & 11 54 22.8113 & $-$02 51 27.8139 & 19.10 & 7.25 \\
\hline
SDSSJ120047.44$+$423744.9 & 6634-56367-554 & 0.3335 & 2.8598 & 12 00 47.4463 & $+$42 37 44.9139 & 18.32 & 26.65 \\
\hline
SDSSJ120708.31$+$354719.6 & 4610-55621-789 & 0.5757 & 2.3012 & 12 07 08.3167 & $+$35 47 19.6774 & 19.57 & 18.51 \\
\hline
SDSSJ121559.87$+$043217.7 & 4749-55633-934 & 0.4847 & 2.8607 & 12 15 59.8718 & $+$04 32 17.7377 & 19.73 & 14.47 \\
\hline
SDSSJ122107.55$+$091612.1 & 5396-55947-145 & 0.4847 & 2.7798 & 12 21 07.5549 & $+$09 16 12.1870 & 19.49 & 7.89 \\
\hline
SDSSJ122440.53$+$222048.7 & 5982-56074-599 & 0.5051 & 2.7236 & 12 24 40.5322 & $+$22 20 48.7335 & 19.57 & 14.69 \\
\hline
SDSSJ122910.34$+$660924.4 & 7120-56720-417 & 0.5086 & 2.4109 & 12 29 10.3491 & $+$66 09 24.4775 & 19.43 & 29.41 \\
\hline
SDSSJ123248.00$+$622451.6 & 7118-56686-839 & 0.5796 & 2.1935 & 12 32 48.0066 & $+$62 24 51.6824 & 19.65 & 30.68 \\
\hline
SDSSJ123516.83$-$014605.4 & 3778-55213-319 & 0.5390 & 2.1920 & 12 35 16.8384 & $-$01 46 05.4520 & 19.45 & 29.72 \\
\hline
SDSSJ124132.92$+$253258.9 & 5984-56337-186 & 0.5372 & 2.7065 & 12 41 32.9224 & $+$25 32 58.9055 & 19.40 & 21.57 \\
\hline
SDSSJ124144.23$+$602441.0 & 6840-56685-377 & 0.5259 & 2.8350 & 12 41 44.2383 & $+$60 24 41.0669 & 19.20 & 19.05 \\
\hline
SDSSJ124300.16$+$262312.5 & 5984-56337-923 & 0.4504 & 2.8910 & 12 43 00.1648 & $+$26 23 12.5652 & 18.78 & 17.27 \\
\hline
SDSSJ125251.16$+$005805.7 & 3850-55575-653 & 0.5399 & 2.4345 & 12 52 51.1670 & $+$00 58 05.7948 & 19.80 & 12.49 \\
\hline
SDSSJ125424.73$+$235639.9 & 5989-56312-735 & 0.6689 & 2.5560 & 12 54 24.7375 & $+$23 56 39.9458 & 19.82 & 17.02 \\
\hline
SDSSJ125820.55$+$451139.1 & 6619-56371-904 & 0.4582 & 2.6481 & 12 58 20.5591 & $+$45 11 39.1837 & 19.42 & 15.43 \\
\hline
SDSSJ125847.45$+$072004.2 & 5417-55978-379 & 0.4700 & 2.5032 & 12 58 47.4573 & $+$07 20 04.2801 & 19.07 & 18.73 \\
\hline
SDSSJ125928.05$+$612722.1 & 6967-56447-575 & 0.4981 & 2.8741 & 12 59 28.0554 & $+$61 27 22.1265 & 18.21 & 13.14 \\
\hline
SDSSJ125941.63$+$024445.7 & 4005-55325-636 & 0.3350 & 2.4306 & 12 59 41.6382 & $+$02 44 45.7256 & 18.21 & 11.78 \\
\hline
SDSSJ130222.99$+$045135.2 & 4758-55682-734 & 0.4993 & 2.8483 & 13 02 22.9980 & $+$04 51 35.2693 & 19.52 & 11.54 \\
\hline
SDSSJ132159.23$+$301023.4 & 6489-56329-661 & 0.4468 & 2.4631 & 13 21 59.2346 & $+$30 10 23.4810 & 19.69 & 10.21 \\
\hline
SDSSJ132525.47$+$393914.5 & 4707-55653-211 & 0.5082 & 2.8280 & 13 25 25.4700 & $+$39 39 14.5486 & 19.81 & 11.33 \\
\hline
SDSSJ132530.49$+$032328.3 & 4761-55633- 55 & 0.4799 & 2.5226 & 13 25 30.4980 & $+$03 23 28.3434 & 19.31 & 12.16 \\
\hline
SDSSJ132808.67$+$615717.6 & 6817-56455-897 & 0.5310 & 2.1957 & 13 28 08.6792 & $+$61 57 17.6358 & 19.63 & 29.19 \\
\hline
SDSSJ134225.98$+$444856.1 & 6628-56366-651 & 0.6208 & 2.1468 & 13 42 25.9827 & $+$44 48 56.1099 & 19.57 & 22.94 \\
\hline
SDSSJ134403.20$+$364645.7 & 3852-55243-491 & 0.4582 & 2.7911 & 13 44 03.2080 & $+$36 46 45.7159 & 19.39 & 26.13 \\
\hline
SDSSJ134408.02$+$162127.9 & 5439-56045-109 & 0.4957 & 2.3610 & 13 44 08.0200 & $+$16 21 27.9890 & 19.28 & 49.16 \\
\hline
SDSSJ134535.17$+$470306.6 & 6749-56370-435 & 0.5260 & 2.6196 & 13 45 35.1782 & $+$47 03 06.6714 & 19.65 & 18.89 \\
\hline
SDSSJ140455.24$+$460020.4 & 6750-56367-171 & 0.4588 & 2.4727 & 14 04 55.2466 & $+$46 00 20.4620 & 19.16 & 9.72 \\
\hline
SDSSJ140639.02$-$020526.4 & 4038-55363-819 & 0.5304 & 2.7589 & 14 06 39.0234 & $-$02 05 26.4441 & 19.73 & 8.94 \\
\hline
SDSSJ141026.39$+$212852.3 & 5896-56047-358 & 0.3135 & 2.9470 & 14 10 26.3965 & $+$21 28 52.3036 & 18.18 & 16.46 \\
\hline
SDSSJ141358.21$+$293240.4 & 6497-56329-271 & 0.4461 & 2.7151 & 14 13 58.2129 & $+$29 32 40.4073 & 19.41 & 17.63 \\
\hline
SDSSJ141613.97$+$301457.0 & 6497-56329-843 & 0.5666 & 2.3858 & 14 16 13.9709 & $+$30 14 57.0543 & 19.50 & 5.82 \\
\hline
SDSSJ141815.72$+$015832.2 & 4030-55634-117 & 0.5559 & 2.1388 & 14 18 15.7288 & $+$01 58 32.2270 & 19.35 & 28.14 \\
\hline
SDSSJ142006.70$+$320348.3 & 3867-55652-544 & 0.7268 & 2.7780 & 14 20 06.7017 & $+$32 03 48.3234 & 19.66 & 12.98 \\
\hline
SDSSJ142516.87$+$142856.5 & 5461-56018-142 & 0.3710 & 2.4583 & 14 25 16.8750 & $+$14 28 56.4956 & 18.46 & 9.16 \\
\hline
SDSSJ142823.28$+$235839.2 & 6014-56072-764 & 0.4660 & 2.7911 & 14 28 23.2874 & $+$23 58 39.2230 & 19.36 & 18.49 \\
\hline
SDSSJ142954.80$+$120235.5 & 5463-56003-121 & 0.5531 & 2.8253 & 14 29 54.8035 & $+$12 02 35.5836 & 19.60 & 25.57 \\
\hline
SDSSJ143133.29$-$004454.1 & 4025-55350-449 & 0.4400 & 2.4480 & 14 31 33.2959 & $-$00 44 54.1727 & 21.07 & 46.76 \\
\hline
SDSSJ143316.32$+$511222.0 & 6718-56398-321 & 0.5323 & 2.5997 & 14 33 16.3257 & $+$51 12 22.0853 & 19.46 & 19.36 \\
\hline
SDSSJ143719.80$+$335056.1 & 3865-55272-411 & 0.4725 & 2.2792 & 14 37 19.8083 & $+$33 50 56.1356 & 18.99 & 7.85 \\
\hline
SDSSJ144021.10$+$200416.9 & 5904-56046-363 & 0.5452 & 2.2529 & 14 40 21.1084 & $+$20 04 16.9359 & 19.91 & 23.85 \\
\hline
SDSSJ144317.83$+$510721.0 & 6725-56390-772 & 0.5501 & 2.9443 & 14 43 17.8381 & $+$51 07 21.0187 & 18.72 & 16.88 \\
\hline
SDSSJ144320.72$+$334212.1 & 3865-55272- 67 & 0.5786 & 2.2016 & 14 43 20.7275 & $+$33 42 12.1014 & 19.56 & 54.20 \\
\hline
SDSSJ144431.48$+$371139.7 & 5173-56046- 85 & 0.4873 & 2.1913 & 14 44 31.4832 & $+$37 11 39.7192 & 19.40 & 17.68 \\
\hline
SDSSJ150114.60$+$304230.8 & 3875-55364-935 & 0.6384 & 2.6498 & 15 01 14.6082 & $+$30 42 30.8537 & 19.44 & 14.96 \\
\hline
SDSSJ150221.28$-$001425.6 & 4016-55632-383 & 0.6272 & 2.2230 & 15 02 21.2805 & $-$00 14 25.6128 & 19.76 & 28.97 \\
\hline
SDSSJ150256.13$+$382017.8 & 5168-56035-693 & 0.4531 & 2.6692 & 15 02 56.1328 & $+$38 20 17.8894 & 18.74 & 19.59 \\
\hline
SDSSJ151017.06$+$193935.8 & 3956-55656-289 & 0.5222 & 2.9298 & 15 10 17.0618 & $+$19 39 35.8621 & 19.58 & 12.96 \\
\hline \hline
\end{tabular}
\end{center}
\end{table*}
\addtocounter{table}{-1}
\begin{table*}[htbp]
\begin{center}
\caption{\textit{Continued}}
\begin{tabular}{l c c c c c c c}
\hline \hline
Target & Plate--MJD--Fiber & $z_{L}$ & $z_{s}$ & R.A. & Decl. & $m_i$ & Ly$\alpha$ Flux \\
\hline
SDSSJ151921.73$+$242649.7 & 3961-55654-899 & 0.4608 & 2.9434 & 15 19 21.7310 & $+$24 26 49.7786 & 18.72 & 9.47 \\
\hline
SDSSJ152105.06$+$254110.0 & 3964-55648- 61 & 0.5987 & 2.4306 & 15 21 05.0684 & $+$25 41 10.0081 & 19.65 & 23.06 \\
\hline
SDSSJ152851.87$+$310233.8 & 4723-56033-445 & 0.7426 & 2.1482 & 15 28 51.8701 & $+$31 02 33.8910 & 19.45 & 15.76 \\
\hline
SDSSJ153635.76$+$024211.9 & 4054-55358-721 & 0.5169 & 2.1774 & 15 36 35.7642 & $+$02 42 11.9110 & 19.27 & 19.96 \\
\hline
SDSSJ153808.49$+$063010.6 & 4885-55735- 15 & 0.5296 & 2.8332 & 15 38 08.4961 & $+$06 30 10.6894 & 19.61 & 17.62 \\
\hline
SDSSJ154204.99$+$401954.6 & 4976-56046-672 & 0.4351 & 2.4976 & 15 42 04.9951 & $+$40 19 54.6945 & 18.78 & 38.28 \\
\hline
SDSSJ154533.14$+$114438.4 & 4886-55737-229 & 0.5087 & 2.2845 & 15 45 33.1494 & $+$11 44 38.4325 & 19.10 & 13.13 \\
\hline
SDSSJ154611.14$+$280542.5 & 3952-55330-944 & 0.3044 & 2.7125 & 15 46 11.1438 & $+$28 05 42.5880 & 17.95 & 17.26 \\
\hline
SDSSJ155458.11$+$264630.7 & 3946-55659-954 & 0.2914 & 2.4440 & 15 54 58.1177 & $+$26 46 30.7333 & 17.79 & 12.73 \\
\hline
SDSSJ155821.52$+$265829.7 & 4725-55711-295 & 0.5788 & 2.3788 & 15 58 21.5222 & $+$26 58 29.7679 & 19.66 & 24.13 \\
\hline
SDSSJ160007.29$+$145112.2 & 3923-55325-932 & 0.3416 & 2.3649 & 16 00 07.2986 & $+$14 51 12.2340 & 18.14 & 12.08 \\
\hline
SDSSJ160324.84$+$203108.4 & 3924-55332-559 & 0.5527 & 2.1657 & 16 03 24.8438 & $+$20 31 08.4723 & 19.76 & 30.28 \\
\hline
SDSSJ160348.28$+$443842.4 & 6034-56103-572 & 0.4823 & 2.6138 & 16 03 48.2849 & $+$44 38 42.4411 & 19.28 & 12.41 \\
\hline
SDSSJ161514.04$+$094650.7 & 5205-56040-279 & 0.4643 & 2.5324 & 16 15 14.0442 & $+$09 46 50.7319 & 19.49 & 12.35 \\
\hline
SDSSJ162432.12$+$524049.6 & 6313-56460-757 & 0.3707 & 2.8857 & 16 24 32.1277 & $+$52 40 49.6765 & 19.45 & 10.99 \\
\hline
SDSSJ165003.84$+$494249.8 & 6315-56181-654 & 0.3889 & 2.6861 & 16 50 03.8489 & $+$49 42 49.8120 & 18.15 & 7.38 \\
\hline
SDSSJ172646.46$+$270324.9 & 5004-55711-565 & 0.5633 & 2.0505 & 17 26 46.4648 & $+$27 03 24.9019 & 19.38 & 31.25 \\
\hline
SDSSJ173225.89$+$264240.3 & 5004-55711-151 & 0.5029 & 2.1957 & 17 32 25.8911 & $+$26 42 40.3294 & 19.93 & 12.43 \\
\hline
SDSSJ212047.63$-$000911.3 & 4192-55469-151 & 0.5470 & 2.6590 & 21 20 47.6367 & $-$00 09 11.3327 & 19.94 & 27.13 \\
\hline
SDSSJ212932.71$+$080751.9 & 4085-55452-895 & 0.4139 & 2.5316 & 21 29 32.7173 & $+$08 07 51.9658 & 18.44 & 20.20 \\
\hline
SDSSJ214843.04$+$044436.1 & 4091-55498-227 & 0.6584 & 2.6963 & 21 48 43.0444 & $+$04 44 36.1066 & 19.85 & 10.03 \\
\hline
SDSSJ215034.63$+$245016.5 & 5960-56097-209 & 0.5039 & 2.2822 & 21 50 34.6362 & $+$24 50 16.5367 & 19.10 & 13.36 \\
\hline
SDSSJ221321.15$+$090241.0 & 5068-55749-829 & 0.6295 & 2.6666 & 22 13 21.1523 & $+$09 02 41.0630 & 19.79 & 15.83 \\
\hline
SDSSJ221952.65$+$104226.6 & 5047-55833-169 & 0.5029 & 2.2230 & 22 19 52.6538 & $+$10 42 26.6000 & 19.29 & 18.09 \\
\hline
SDSSJ222151.54$+$072400.2 & 5069-56211-141 & 0.7507 & 2.5072 & 22 21 51.5479 & $+$07 24 00.2355 & 19.64 & 13.67 \\
\hline
SDSSJ222234.29$+$200711.6 & 5023-55858-985 & 0.6448 & 2.4528 & 22 22 34.2920 & $+$20 07 11.6048 & 19.88 & 11.16 \\
\hline
SDSSJ223301.06$+$100921.6 & 5053-56213-894 & 0.5684 & 2.4456 & 22 33 01.0693 & $+$10 09 21.6623 & 19.71 & 28.09 \\
\hline
SDSSJ223333.16$+$272037.5 & 6297-56218-817 & 0.5324 & 2.6963 & 22 33 33.1641 & $+$27 20 37.5412 & 19.86 & 10.90 \\
\hline
SDSSJ223622.28$+$272332.8 & 6297-56218-923 & 0.5299 & 2.8474 & 22 36 22.2803 & $+$27 23 32.8281 & 19.90 & 14.63 \\
\hline
SDSSJ223843.98$+$151959.4 & 5038-56235- 77 & 0.5717 & 2.3057 & 22 38 43.9819 & $+$15 19 59.4942 & 20.04 & 18.58 \\
\hline
SDSSJ224704.56$+$162549.5 & 5033-56244-339 & 0.2927 & 2.3936 & 22 47 04.5630 & $+$16 25 49.5117 & 18.80 & 72.74 \\
\hline
SDSSJ224746.60$-$024237.9 & 4364-55855-469 & 0.4176 & 2.7720 & 22 47 46.6040 & $-$02 42 37.8978 & 19.35 & 11.56 \\
\hline
SDSSJ230217.65$+$302738.2 & 6506-56564-729 & 0.5612 & 2.8060 & 23 02 17.6514 & $+$30 27 38.2901 & 19.78 & 13.51 \\
\hline
SDSSJ230827.85$+$311814.9 & 6504-56540-258 & 0.4096 & 2.6515 & 23 08 27.8540 & $+$31 18 14.9936 & 18.69 & 378.79 \\
\hline
SDSSJ230917.33$+$072438.6 & 6168-56187-395 & 0.5606 & 2.2883 & 23 09 17.3364 & $+$07 24 38.6568 & 19.48 & 18.67 \\
\hline
SDSSJ231418.03$-$011027.0 & 4209-55478-171 & 0.5221 & 2.2679 & 23 14 18.0322 & $-$01 10 27.0288 & 19.85 & 20.38 \\
\hline
SDSSJ231731.62$+$115450.9 & 6147-56239-163 & 0.5448 & 2.6389 & 23 17 31.6260 & $+$11 54 50.9168 & 19.62 & 38.16 \\
\hline
SDSSJ232231.78$+$144340.7 & 6143-56267-511 & 0.4522 & 2.4062 & 23 22 31.7871 & $+$14 43 40.7233 & 18.93 & 13.05 \\
\hline
SDSSJ233240.44$+$150308.2 & 6137-56270-331 & 0.5096 & 2.3695 & 23 32 40.4443 & $+$15 03 08.1958 & 19.73 & 21.74 \\
\hline
SDSSJ233311.11$+$022310.9 & 4282-55507-607 & 0.4716 & 2.2529 & 23 33 11.1182 & $+$02 23 10.9309 & 19.92 & 24.45 \\
\hline
SDSSJ234617.29$+$090931.4 & 6160-56190-854 & 0.1817 & 2.7125 & 23 46 17.2925 & $+$09 09 31.4367 & 17.35 & 27.96 \\
\hline
SDSSJ234653.04$-$032113.7 & 4356-55829-185 & 0.6154 & 2.7841 & 23 46 53.0420 & $-$03 21 13.6990 & 19.97 & 19.92 \\
\hline
SDSSJ235212.72$+$214526.1 & 6521-56537-324 & 0.5133 & 2.9180 & 23 52 12.7222 & $+$21 45 26.1749 & 19.74 & 14.22 \\
\hline
SDSSJ235356.11$-$083238.8 & 7166-56602-783 & 0.4790 & 2.6038 & 23 53 56.1108 & $-$08 32 38.8589 & 19.27 & 13.98 \\
\hline
SDSSJ235508.04$-$052826.7 & 7033-56565-257 & 0.5572 & 2.5178 & 23 55 08.0493 & $-$05 28 26.7535 & 19.71 & 8.76 \\
\hline
SDSSJ235525.24$-$052138.6 & 7033-56565-181 & 0.4917 & 2.2431 & 23 55 25.2466 & $-$05 21 38.6252 & 19.28 & 13.67 \\
\hline \hline
\end{tabular}
\end{center}
\end{table*}

\end{document}